\documentclass[12pt]{article}

\usepackage{jheppub} 

\usepackage{amsmath,amsfonts}
\usepackage{amssymb}
\usepackage{fmtcount}

\usepackage{braket}

\usepackage{slashed}
\renewcommand{\sl}{\slashed}

\newcommand{\be}{\begin{equation}}
\newcommand{\ee}{\end{equation}}
\newcommand{\bea}{\begin{eqnarray}}
\newcommand{\eea}{\end{eqnarray}}
\newcommand{\Tr}{\operatorname{Tr}}
\newcommand{\tr}{\operatorname{tr}}

\newcommand{\nn}{\nonumber}

\newcommand{\gen}{\mathfrak t}

\newcommand{\ben}{\begin{enumerate}}
\newcommand{\een}{\end{enumerate}}

\newcommand{\s}{\sigma}
\renewcommand{\t}{\tau}
\renewcommand{\r}{\varrho}

\newcommand{\phih}{\Phi}
\newcommand{\Ah}{{\mathcal  A}}
\newcommand{\Y}{G}

\newcommand{\cl}[1]{{\left[#1\right]}}

\renewcommand{\cl}[1]{{\left[#1\right]}}

\title{New Covariant Feynman Rules for Effective Field Theories}
\author[a]{Gero von Gersdorff}
\author[a]{and Kevin Santos}
\affiliation[a]{Pontificia Universidade Católica do Rio de Janeiro\\ 
Rua Marquês de São Vicente 225, Rio de Janeiro, Brazil}

\emailAdd{gersdorff@puc-rio.br}
\emailAdd{skevin@aluno.puc-rio.br}

\abstract{We provide a new and completely general formalism to compute the effective field theory matching contributions from integrating out massive fields in a manifestly gauge covariant way, at any desired loop order. The formalism is based on old ideas such as the background field method and the heat kernel, however we add some crucial new ingredients that greatly improve the simplicity and general applicability of the approach. We formulate our method in terms of Feynman rules, the resulting effective action is expressed in terms of local heat kernel coefficients.
We also provide as supplementary material a mathematica code that facilitates the computation of these coefficients.}

\begin{document}

\maketitle
\flushbottom

\section{Introduction}

Effective field theories (EFT's) constitute one of the most important tools in high energy physics, in particular in the perturbative regime.
The fact that a theory with heavy particles can be represented in the infrared (IR) by a simpler one with only 
light degrees of freedom and a series of higher dimensional operators greatly simplifies the phenomenological analysis of the theory, providing an accurate way of obtaining low-energy observables directly related to experiment (see ref.~\cite{Manohar:2018aog} for a review). The problem is broken down into two simpler ones, the matching of the full UV theory to the EFT (the calculation of the  coefficients of the higher dimensional operators), and the calculation of IR observables directly in the EFT, employing the renormalization group (RG) for improved precision.

In recent years, a lot of effort has gone into studying the Standard Model effective field theory (SMEFT). The full list of dimension-six operators has been obtained in ref.~\cite{Grzadkowski:2010es}, and their complete RG equations have been computed in refs.~\cite{Jenkins:2013zja,Jenkins:2013wua,Alonso:2013hga}.
Countless phenomenological studies exist relating the Wilson coefficients to experimental data, putting limits on their values which in turn sharpen our understanding of possible extensions of the Standard Model (SM).
This very model independent analysis  eventually has to be completed by the first step mentioned above, the so-called matching.

While matching is in general quite straightforward at the tree-level, in many perturbative UV theories not all operators are generated in this leading order, and loop calculations are necessary. Moreover, some of these operators are directly connected to 
very precise experimental data, so even loop-suppressed Wilson coefficients can be of great value for putting bounds on possible UV theories. Examples include dipole operators mediating $\mu\to e\gamma$ transitions as well as contributing to electric and magnetic dipole moments of light particles.

Loop calculations are traditionally performed in a non-gauge covariant approach, breaking up covariant derivatives into partial derivatives and gauge fields. This allows for a simple evaluation of the diagrams in momentum space, but manifest gauge invariance becomes hidden and is ensured only by intricate relations between different diagrams (Ward identities).  
Diagrams containing a certain number of external gauge fields have to be combined with diagrams with fewer gauge boson lines but more powers of external momenta to  assemble into gauge invariant operators. Usually a large number of such diagrams has to be computed in order to obtain all possible operators of a desired dimension.

In contrast to this canonical non-covariant perturbation theory, a covariant alternative has been developed a long time ago \cite{schwinger1951,dewitt1965dynamical,DeWitt:1967ub,Gilkey:1975iq,Barvinsky:1985an,Avramidi:1990ug,Avramidi:1990je}. The first manifestly covariant formalism appeared in the seminal work of Schwinger \cite{schwinger1951} (see ref.~\cite{Schwartz:2014sze} for a pedagogical introduction), and was later further pioneered by DeWitt \cite{dewitt1965dynamical,DeWitt:1967ub}, who also recognized its applicability to gravity.
The approach is based on the background field method in which the effective action is obtained by integrating the quantum fluctuations over a classical background. Within this approach it is possible to maintain manifest background field covariance throughout all calculations.
 Schwinger and deWitt  advocated the use of the so-called heat kernel (HK), allowing for an efficient calculation of the one-loop effective action of bosonic background fields (including scalars, gauge fields, and gravity). 
The Schwinger-DeWitt technique was considerably generalized for arbitrary one-loop graphs (including fermionic background fields as well as different spins and masses in the loop) in \cite{Barvinsky:1985an}.  
Most notably, the covariant HK method has been applied to chiral perturbation theory \cite{Gasser:1983yg} and to the calculation of anomalies in various dimensions \cite{Fujikawa:1979ay,Fujikawa:1980eg,Ball:1988xg,vonGersdorff:2003dt,vonGersdorff:2006nt}.  
The method has also been used in the context of extra dimensions \cite{Hoover:2005uf,Barvinsky:2005qi,vonGersdorff:2008df}.
The background field method for higher loops was developed in refs.~\cite{Abbott:1980hw, Boulware:1980av,tHooft:1975uxh}, and has been used in its covariant form to calculate $\beta$ functions for gauge theories at various loop orders \cite{Duff:1975ue,Batalin:1976uv,Batalin:1978gt,Bornsen:2002hh} by considering covariantly constant background field strengths. 
The covariant approach typically reduces by a lot the number of independent diagrams to be computed.

In recent years, a number of authors returned to the idea of  covariant perturbation theory 
\cite{Henning:2014wua,Drozd:2015rsp,delAguila:2016zcb,Henning:2016lyp,Zhang:2016pja,FuentesMartin:2016uol,Ellis:2017jns,Kramer:2019fwz,Ellis:2020ivx,Angelescu:2020yzf,Cohen:2020fcu,Dittmaier:2021fls}.
The new formalism replaces the  HK methods with some  
functional trace techniques \cite{Gaillard:1985uh} that appear to be designed for one-loop determinants only.
This revival has lead to the development of  comprehensive software packages that automate these calculations at the one-loop level \cite{DasBakshi:2018vni,Cohen:2020qvb,Fuentes-Martin:2020udw,Uhlrich:2021ded,Carmona:2021xtq,Fuentes-Martin:2022jrf}.
 
However, despite the decent amount of literature and recent progress on covariant perturbation theory, a completely universal treatment valid to any loop order is still missing. The purpose of the present work is to provide such a simple and universally applicable framework. 
Our main new idea is to factorize each Feynman diagram into an "$n$-point function" $\Gamma(x_1\dots x_n$) that is gauge invariant  at any of the vertices\footnote{This also includes certain bilinear vertices, in particular one-loop graphs will in general have $n\neq 0$ in our formulation.} $x_i$ of the diagram,  
and a momentum space loop integral which depends on $n$ external momenta $I(p_1\dots p_n)$. 
The contribution to the effective Lagrangian is then simply given by $I(i\partial_1\dots i\partial_n)\Gamma(x_1\dots x_n)|_{x_i=x}$, which can be systematically calculated in an expansion of the background fields and their covariant derivatives, making essential use of HK techniques.
This simple structure of loop graphs is quite powerful but to the best of our knowledge has never been exploited in the literature. 
Our method has no restrictions whatsoever for the types of interactions, loop order, types of background fields, loop masses etc.\footnote{We however restrict ourselves to flat gravitational backgrounds for clarity.}
We illustrate our formalism with examples at one and two loops.

This paper is organized as follows. 
In sec.~\ref{sec:hk} we review the concept of the HK representation of the propagators. 
In sec.~\ref{sec:rules} we present and proof our generalized Feynman rules. In sec.~\ref{sec:examples} we illustrate our formalism with examples at one and two-loop order. In appendix \ref{sec:fluc} we review the background field method and write explicitly the field dependent masses and vertices for gauge theories.
In app.~\ref{sec:b2n} we review the techniques for the calculations of the HK coefficients and tabulate the leading ones.
In app.~\ref{sec:integrals}, we summarize some basic integrals over Schwinger parameters.
In the supplementary material to this article we provide a short mathematica notebook that computes HK coefficients with arbitrary number of covariant derivatives.

\section{Heat kernel representation of propagators}
\label{sec:hk}

\subsection{Scalars}

We work in Lorentzian signature. 
Consider the background-field dependent Feynman propagator for a scalar field in an irreducible representation 
of the gauge and (if present) global symmetry group:
\be
G_s(m,X)=\frac{-i}{D^2+m^2+X-i\epsilon}\,.
\ee
The mass $m$ is field-independent and  proportional to the identity, it may be zero. In contrast, the mass $X$ is field dependent, but, since we consider irreducible representations,  $X$ does not mix different representations. 
Representation-mixing field dependent masses will be treated as mass insertions (bilinear vertices). The gauge fixing procedure may generate contribution to $X$, see eq.~(\ref{eq:Xs}).
In the following, we will  suppress the $i\epsilon$ terms.

The heat kernel (HK) representation of the propagator is defined as \cite{Fock:1937dy,Nambu:1950rs,schwinger1951}
\be
G_s(m,X)=\int_0^\infty dt\ K_s(t,m,X)\,,\qquad
K_s(t,m,X)\equiv e^{-i(D^2+m^2+X)t}\,.
\label{eq:HKrep}
\ee
The HK $K_s(t,m,X)$ satisfies the Schrödinger equation\footnote{The name "heat kernel" originates from the Euclidean analogue which satisfies the heat equation.}
\be
i\partial_t K_s(t,m,X)=(D^2+m^2+X) K_s(t,m,X)\,,\qquad K_s(0,m,X)=1\,.
\label{eq:SE}
\ee
Notice that for the case of a scalar, $X$ is Hermitian. However, for nonzero spin (see below), there is a contribution to the effective mass matrix given by the term $-S^{\mu\nu}F_{\mu\nu}$, where $S_{\mu\nu}$ are the generators for Lorentz transformations which are non-Hermitian in non-Euclidean spacetime signature.

Defining $K_s(t,m,X;x,y)\equiv\braket{x|K_s(t,m,X)|y}$, one now makes the ansatz
\be
K_s(t,m,X;x,y)=K_0(t,m;x-y){B(t,X;x,y)}\,,
\ee
where
\be
K_0(t,m;x-y)\equiv\int \frac{d^dk}{(4\pi)^d} e^{i(k^2-m^2)t}e^{-ik(x-y)}=i(4 \pi i t)^{-\frac{d}{2}}e^{-i\frac{(x-y)^2}{4t}-im^2t}\,,
\label{eq:K0def}
\ee
is the free HK (that is, the HK of a gauge singlet with vanishing $X$). The function $B(t)$ 
has a regular expansion around $t=0$:
\be
B(t,X;x,y)\equiv \sum_{n\geq 0} b_{2n}(X;x,y)\frac{t^n}{n!}\,.
\label{eq:hkexp}
\ee
The $b_{2n}$ are known as the {\em heat kernel coefficients}.   
In particular, the coefficient $b_0$, which is independent of $X$, is equal to the Wilson line along the straight line segment connecting $x$ to $y$ (see app.~\ref{sec:b2n}).
The higher coefficients have no closed-form expressions, however all we will need are their various covariant derivatives evaluated at coinciding arguments ($x=y$), which we will refer to as the local heat kernel coefficients (LHKC's). These objects are local polynomials in the background fields and can be computed systematically \cite{dewitt1965dynamical,DeWitt:1967ub,Gilkey:1975iq,Avramidi:1990je,Vassilevich:2003xt}.
If the  field transforms in the $r$ representation, $B$ transforms as the $r$ representation at $x$ (that is, from the left) and as the conjugate $\bar r$ representation at $y$ (from the right), since $ G^\alpha_{\ \beta}(x,y)=\braket{\phih^\alpha(x)\phih_\beta^\dagger(y)}$. $K_0$  transforms as a singlet.
We will be using the HK in a "mixed position-momentum representation" 
\be
K_s(t,m,X;x,y)=\int \frac{d^dk}{(2\pi)^d}\,e^{-ik(x-y)}\biggl( e^{i(k^2-m^2)t}B(t,X;x,y)\biggr)\,.
\label{eq:scalarHK}
\ee

Sometimes an interaction contains a covariant derivative acting on a fluctuation field (this happens, for instance, in a gauge theory with scalars, see app.~\ref{sec:fluc}). We will associate this derivative to the propagator directly, that is we also will need propagators
\be
\braket{D_\mu\phih(x)\phih^\dagger(y)}=D^x_\mu G_s(m,X;x,y)\,,
\ee
and similar terms with a derivative acting on $y$ as well as both arguments. Such terms are easy to deal with in our formalism as $D_\mu G_s=\int dt\ D_\mu K_s(t)$, and
\be
D_\mu^x K_s(t,m,X;x,y)=\int \frac{d^dk}{(2\pi)^d}\,e^{-ik(x-y)}\biggl( e^{i(k^2-m^2)t}(-ik_\mu+D_\mu^x)B(t,X;x,y)\biggr)\,,
\label{eq:propder}
\ee
etc. 

One may evaluate explicitly the integral over the Schwinger parameter $t$ to obtain the propagator in terms of the HK coefficients:
\be
G_s(m,X;x,y)=\int \frac{d^dk}{(2\pi)^d}\, e^{-ik(x-y)} \left(\sum_{n=0}^\infty\left(\frac{i }{k^2-m^2}\right)^{n+1} b_{2n}(X;x,y)\right)\,.
\label{eq:prophk}
\ee
Even though the Feynman rules can be given equally well in terms of this propagator instead of the HK, we find it more practical to perform the integration over the Schwinger parameters last (in particular, after the momentum integration), so eq.~(\ref{eq:prophk}) is not relevant for us in practice. 
There is however one important observation that can be made from eq.~(\ref{eq:prophk}). Suppose $X$ is non-Hermitian, which, as we already mentioned, can occur for fields of nonzero spin. The representation of the propagator in terms of the HK in eq.~(\ref{eq:HKrep}) is not well defined for arbitrary complex eigenvalues. However, all we are interested in is the local part of the effective action, which is implied in the small-$t$ expansion of eq.~(\ref{eq:hkexp}). If we {\em define} our propagators  in terms of this expanded HK, we obtain the well defined object in eq.~(\ref{eq:prophk}).

\subsection{Spin-$\frac{1}{2}$ fermions}
We start by writing the fermion propagator in terms of the scalar one:
\footnote{In the presence of a field dependent mass $Y+i\gamma_5Z$, we can instead write
\be 
G(m,Y,Z)=
i\left(i\sl D-m-Y-i\gamma_5 Z\right)^{-1}
=(-i)(i\sl D+m+Y-i\gamma_5 Z)\left(D^2+m^2+X\right)^{-1}
\ee
with $X= -S^{\mu\nu}F_{\mu\nu}-i\sl DY-\gamma_5\sl D Z+2mY+Y^2+Z^2$. In most applications $Y=Z=0$.} 
\be
G_{\!f}(m)=
\frac{i}{i\sl D-m}
=(i\sl D+m)\frac{-i}{D^2+m^2+X_{\!f}}\,,
\qquad
X_{\!f}\equiv -S^{\mu\nu}F^a_{\mu\nu}\gen^a
\,,
\ee
where $S^{\mu\nu}=\frac{i}{4}[\gamma^\mu,\gamma^\nu]$ is the 
Lorentz generator for spin $\frac{1}{2}$.
We then define the corresponding HK
\be
K_{\!f}(t,m)\equiv (i\sl D+m)e^{-i(D^2+m^2+X_{\!f})t }\,,
\ee
such that as previously $G_{\!f}=\int dt\, K_{\!f}(t)$.
The HK  in its mixed representation that we will be using in our Feynman rules is thus
\be
K_{\!f}(t,m;x,y)
=\int \frac{d^dk}{(2\pi)^d}e^{-ik(x-y)}\biggl(e^{i(k^2-m^2)t}(\sl k+m+i\sl D^x)B(t,X_{\!f};x,y)\biggr)\,.
\label{eq:fermionHK}
\ee
We notice that the same function $B$ appears on the right hand side, albeit with a different argument $X$.

\subsection{Gauge fields}
\label{sec:gauge}

Finally, consider the gauge sector. 
It is easiest to use Feynman gauge $\xi=1$ to avoid dealing with non-minimal operators.\footnote{See ref.~\cite{Barvinsky:1985an} for the HK with non-minimal operators.} Then from the terms quadratic in the fluctuations detailed in app.~\ref{sec:fluc} one can read off the propagator 
\be
G_v=\frac{+i}{D^2+X_v-i\epsilon}\,,
\qquad (X_v)^{a\,\mu}_{\ b\,\nu}\equiv -2f^{a}_{\ bc}F^{c\,\mu}_{\ \  \nu}
+g^2\phi^\dagger \{\gen^a,\gen_b\}\phi\delta^\mu_{\ \nu}\,.
\ee
Like in the spin-$\frac{1}{2}$ case, the term in $X_v$ proportional to the field strength can be understood as $-S^{\mu\nu}F^a_{\mu\nu}\gen^a$, with the spin-1 Lorentz generator 
$(S^{\mu\nu})_{\alpha\beta}=i(\delta^\mu_\alpha\delta^\nu_\beta-\delta^\mu_\beta\delta^\nu_\alpha)$ and the adjoint generator $(\gen^a)^b_{\ c}=-if^{ab}_{\ \ c}$.
The gauge HK can be decomposed as
\be
K_{v}(t;x,y)=\int \frac{d^dk}{(2\pi)^d}\,e^{-ik(x-y)}\biggl(-e^{i(k^2-m^2)t}B(t,X_v;x,y)\biggr)\,.
\ee
The gauge and Lorentz indices  on $G_v$, $K_v$ and $B$ are suppressed here for clarity. 
Notice that in the limit of vanishing background fields we obtain $B^{a\,\mu}_{\ \ b\,\nu}=\delta^a_b\delta^\mu_\nu$, recovering the usual propagator in the Feynman gauge, including the correct negative sign.

The HK for the ghost is given by the scalar one, with the field dependent mass 
\be
(X_{g})^{a}_{\ b}=g^2\, \phi^\dagger \{\gen^a, \gen_b\}\phi\,,
\ee
that is of the same structure as the one for the gauge fields.

\section{Feynman rules}
\label{sec:rules}

\begin{table}
\begin{center}
\begin{tabular}{|c|}
\hline
\\
\includegraphics*[width=6 cm]{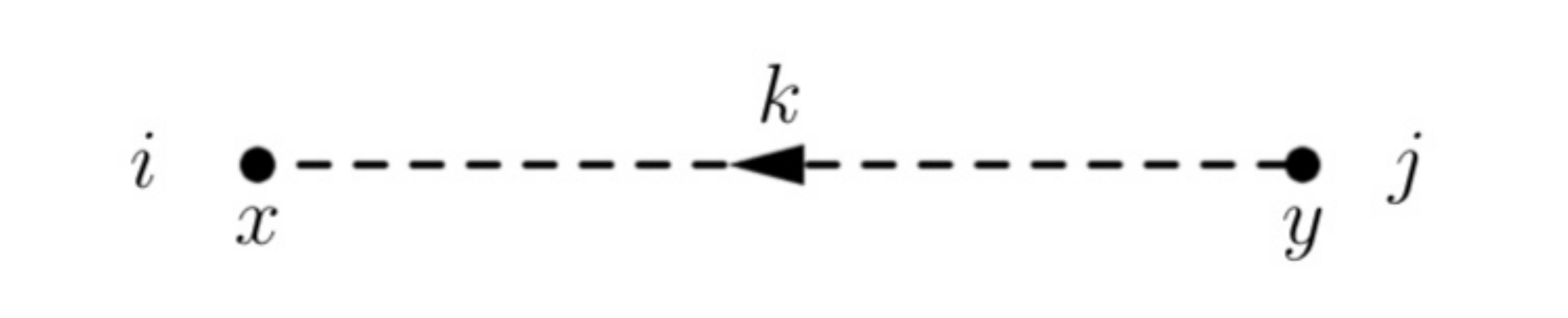}\\
	 $e^{i(k^2-m^2)t}\{B(t,X_s;x,y)\}^i_{\ j}$\\
\\
\hline
\\	 
\includegraphics*[width=6 cm]{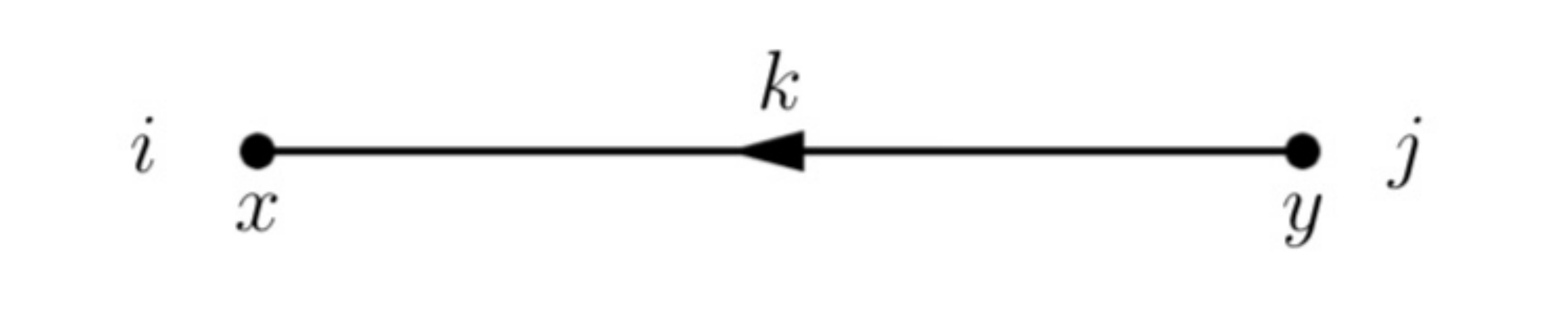}\\
$e^{i(k^2-m^2)t}(\sl k+m+i\sl D^x)\{B(t,X_f;x,y)\}^i_{\ j}$\\
\\
\hline
\\
\includegraphics*[width=6 cm]{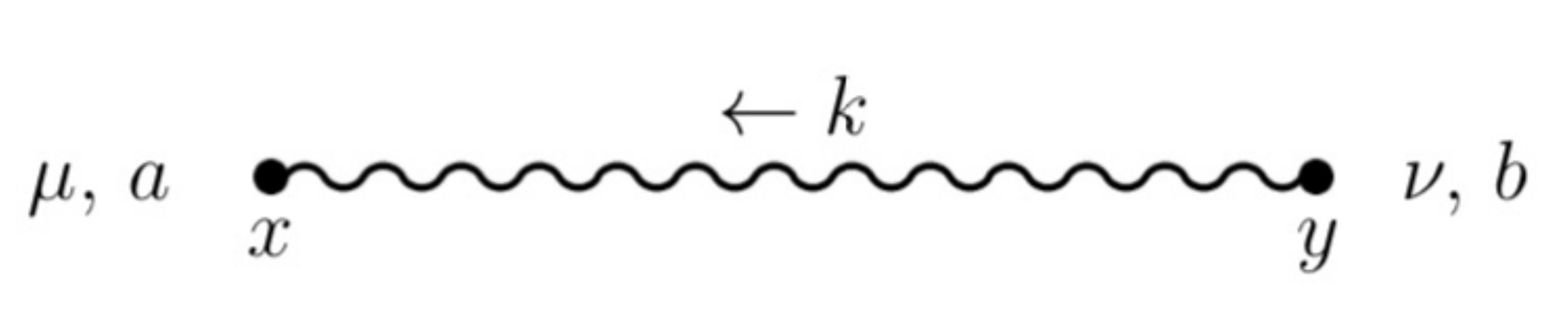}
\\
$-e^{ik^2 t}\left\{B(t,X_v;x,y)\right\}^{a\,\mu}_{\ \ b\, \nu}$\\	
\\
\hline	
\end{tabular}
\end{center}
\caption{Feynman rules for propagators.}
\label{tab:props}
\end{table}

\subsection{Summary of the rules}
\label{sec:summary}

Let us first give a concise list of rules which can be directly applied to any given problem. We will provide a proof for these rules in sec.~\ref{sec:proof}.
\ben
\item Draw all diagrams with with no external legs.
\label{fr:diag}
\item 
\label{fr:prop}
For each  propagator write the expressions given in tab.~\ref{tab:props}.
For propagators of derivatives of fields add additional factors of $(-ik_\mu+D_\mu^x)$ and $(ik_\mu+D_\mu^y)$ as explained in eq.~(\ref{eq:propder}).
The expressions for the fermion propagator and the derivative scalar propagator assumes that the momentum $k$ flows parallel to particle-number.


\item
\label{fr:vert}
Write the field dependent couplings, including the usual factor of $i$ for each of them.

\item
\label{fr:extmom}
To each vertex, associate an ingoing momentum $p_i$, and impose momentum conservation at every vertex.
  Perform the loop momentum integration in $d$ dimensions.
\item
\label{fr:expmom}
Expand in the external momenta $p_i$ to the desired order, 
replace the momenta $p^{i}$ by the partial derivatives
\be
p_\mu^i\to i\frac{\partial}{\partial x_\mu^i}
\ee 
and take the  limits $x_i\to x$ everywhere.
\item
\label{fr:final}
Integrate over the Schwinger parameters, remove all poles via $\overline {MS}$, and take the limit $d\to 4$. The result, multiplied by $-i$, is the EFT matching contribution to the Lagrangian.
\een

If there is a field dependent mass that does not mix representations (i.e., that commutes with the field-independent mass matrix), one may wish to include it into the propagators and heat kernels. Field dependent masses that do mix representations need to be included as mass insertions (bilinear vertices). 

The partial derivatives appearing in step \ref{fr:expmom} always act on a total gauge singlet, essentially the product of all the $B$-factors and field-dependent couplings (see sec.~\ref{sec:proof}). Once these derivatives are distributed over the 
individual (non gauge singlet) factors, covariant derivatives will reappear.

Observe the presence of the coincidence limits $x_i\to x$ in step \ref{fr:expmom}. This means that, as claimed in section \ref{sec:hk},
only the LHKCs, i.e., the coincidence limits of coefficients $b_{2n}$ (and their covariant derivatives) appear.
 A technical simplification that we already would like to point out is that any string of partial derivatives resulting 
from the expansion in external momenta is automatically symmetric (since partial derivatives commute)
\be
\partial_{\mu_1}\partial_{\mu_2}\dots\partial_{\mu_n}=
\partial_{(\mu_1}\partial_{\mu_2}\dots\partial_{\mu_n)}
\label{eq:sympartial}
\ee 
where the parenthesis on the index denotes symmetrization with strength one. This symmetrization will thus appear also in the resulting  covariant derivatives, which greatly simplifies the calculation. For instance, symmetrized covariant derivatives of the ubiquitous coefficient $b_0$ are zero, see app.~\ref{sec:b2n}.

A last comment concerns one-loop diagrams with zero vertices, that is with one single massive particle in the loop. This case is somewhat special since it involves the log of the propagator, but is also the most studied one in the literature \cite{Vassilevich:2003xt}. For bosons one needs the functional trace identity 
\be
\Tr\log (D^2+X+m^2)=-\int_0^\infty \frac{dt}{t} \Tr e^{-i(D^2+X+m^2)t}
\ee
and for Dirac fermions 
\be
\Tr\log (i\sl D-m)=-\frac{1}{2}\int_0^\infty \frac{dt}{t} \Tr e^{-i(D^2-F_{\mu\nu}S^{\mu\nu}+m^2)t}
\ee
These traces are straightforwardly evaluated in terms of traces over the HK coefficients $b_{2n}(x,x)$.

\subsection{Proof of the rules}
\label{sec:proof}

We now proceed to prove the rules given in sec.~\ref{sec:summary}.

Rule \ref{fr:diag}
just states that we are calculating the effective action in the background field method. Lines correspond to propagators of fluctuations and not to background fields, which are hidden in the lines and vertices.

Rule \ref{fr:prop} is evident from the HK representation of the propagators $G=\int dt K(t)$ and the explicit expressions for the HK in eq.~(\ref{eq:scalarHK}) and eq.~(\ref{eq:fermionHK}).
Notice that we are not writing the factors of $e^{ikx}$, they will be combined into momentum conserving delta functions similar to the standard (non-covariant) Feynman rules (see below). We also delay the integrations over the momenta and Schwinger parameters.

Rule \ref{fr:vert} is self-evident.

To show rules \ref{fr:extmom} and \ref{fr:expmom}, let us examine the possible spacetime dependence for a given vertex at $x$ resulting from the rules so far.
For instance, consider a bilinear interaction
\be
\mathcal L_{\rm int}=C_{\alpha\beta}(x) \Phi^{\alpha}(x)\Psi^\beta(x)\,,
\ee
where $\Phi$ and $\Psi$ are two fluctuation fields (of possibly different spins), and $C$ is a background-field dependent coupling. We suppress possible Lorentz indices and only display the gauge indices $\alpha$ and $\beta$, belonging in general to different irreducible representations.
Such an interaction will give rise to factors of the kind 
\be
iC_{\alpha\beta}(x)B^{\alpha,\, \cdot\, }(t;x,\, \cdot\, )B^{\beta,\, \cdot\, }(t';x,\, \cdot\, )\,.
\label{eq:vertex}
\ee
This expression is gauge invariant at $x$, the gauge indices $\alpha, \beta$  are contracted in a gauge-invariant way in the same way as in the Lagrangian. 
For this reason, the product of all propagator $B$ factors and couplings $C$ becomes a {\em complete gauge singlet} that we call
\be
\Gamma(t_1,\dots  t_P,x_1,\dots x_V)\equiv \prod_{i=1}^P B_i(t_i)
\prod_{\ell=1}^V iC_\ell\,, 
\label{eq:Btens}
\ee
where $P$ denotes the number of propagators and $V$ the number of vertices. 
Diagrams with only scalar propagators become (suppressing the $t_i$ integrations for clarity)
\be
\int \prod_{i=1}^P \frac{d^dk_i}{(2\pi)^d}e^{i(k_i^2-m_i^2)t_i}\prod_{\ell=1}^V d^dp_\ell \delta(p_\ell-P_\ell)
\tilde \Gamma(t_1,\dots  t_P,p_1,\dots p_V)\,,
\ee
where $P_\ell$ is the total momentum exiting at the $\ell^{th}$ vertex {in the propagators}.
We have Fourier-transformed the "$n$-point function" $\Gamma$ and performed the integration over the $x_i$ to generate momentum-conserving $\delta$ functions.\footnote{Our convention is that factors of $e^{ikx}$ ($e^{-ikx}$) correspond to momentum $k$ exiting (entering) the vertex at $x$.}
This Fourier transform of $\Gamma$ introduces $V$ new "external" ingoing momenta $p_\ell$, one for each vertex.
The generalization to fermion  propagators and derivative propagators is straightforward: in this case the objects $\Gamma$ in eq.~(\ref{eq:Btens})   may  contain additional derivatives (from terms such as $\sl D$ and $D_\mu$ in eqns.~(\ref{eq:fermionHK}) and (\ref{eq:propder})), and one gains additional factors of propagator momenta  $\sl k$ and $k_\mu$.
These details however are not important for the following arguments. Using momentum conservation we get for our diagram
\be
\int \prod_{\ell=1}^V \frac{d^dp_\ell}{(2\pi)^d} 
\left[(2\pi)^d\delta\bigl(\textstyle\sum_{\ell=1}^V p_\ell\bigr) 
\right]
\tilde \Gamma(t_1,\dots  t_P,p_1,\dots p_V)
\displaystyle \prod_{r=1}^L\frac{d^dq_r}{(2\pi)^d} \prod_{i=1}^P e^{i(k_i^2-m_i^2)t_i}\,,
\ee
where $L$ is the number of loops, and the propagator momenta $k_i$ are now to be interpreted as the corresponding linear combinations of the 
loop momenta $q_r$ and external momenta $p_\ell$ as dictated by momentum conservation. At this time, we can perform the integral over the loop momenta which we call
\be
I(t_1\dots t_P,p_1\dots p_V)\equiv \prod_{r=1}^L\frac{d^dq_r}{(2\pi)^d} \prod_{i=1}^P e^{i(k_i^2-m_i^2)t_i}\,,
\ee
such that (restoring the $t_i$ integrations):
\be
i S_{\rm eff}\supset 
\int \prod_{\ell=1}^V \frac{d^dp_\ell}{(2\pi)^d} 
\left[(2\pi)^d\delta\bigl(\textstyle\sum_{\ell=1}^V p_\ell\bigr) \right]
\displaystyle
\int \prod_{i=1}^P dt_i\
I(t_1\dots t_P,p_1\dots p_V)
\tilde \Gamma(t_1,\dots  t_P,p_1,\dots p_V)\,.
\label{eq:Seffmom}
\ee
Transforming back to position space shows that this is indeed  local upon expansion of $I$:
\be
i S_{\rm eff}\supset \int d^d x \int \prod_{i=1}^P dt_i\ I(t_1\dots t_P,i\partial_1, \dots i\partial_V) \Gamma(t_1,\dots  t_P,x_1,\dots x_V)|_{x_\ell\to x}\,.
\label{eq:FRfinal}
\ee
We note that the integral $I$ is regular at vanishing external momenta, such that a Taylor expansion around zero external momenta is well defined. Eq.~(\ref{eq:FRfinal}) is precisely what is stated in rule \ref{fr:expmom}, including the coincidence limit $x_\ell\to x$.

Eq.~(\ref{eq:FRfinal}) is known as the effective action of the full theory.
To really obtain the  matching contribution, we need to subtract the EFT contribution to the same operators. However, provided that the EFT does no longer contain any massive fields, all EFT integrals are scaleless, and hence vanish in dimensional regularization. For logarithmically divergent integrals, this is due to a cancellation of UV and IR poles. Following the  pedagogical discussion of ref.~\cite{Manohar:2018aog}, we make this cancellation visible by defining $\epsilon_{\rm UV}\equiv \epsilon$, and $\epsilon_{\rm IR}\equiv \epsilon$, and write schematically the result of the EFT contribution as (at one-loop for definiteness)
\be
I_{\rm EFT}+I_{\rm EFT,c.t.}
	=A\left(\frac{1}{\epsilon_{\rm UV}}-\frac{1}{\epsilon_{\rm IR}}\right)-\frac{A}{\epsilon_{\rm UV}}=-\frac{A}{\epsilon_{\rm IR}}
\ee
where "c.t." stands for counterterm and $A$ is some coefficient. Thus, the net effect of the EFT contribution  are some poles in $\epsilon_{\rm IR}$.  
The same poles must occur in the full theory calculation as the two theories by construction reproduce the same IR physics.\footnote{Indeed, 
the HK expansion and the expansion in the external momenta create IR divergences in $d=4$.} Thus, in dimensional regularization, the EFT contribution's only effect is to  cancel the IR divergences of the full theory (of course, any UV divergences of the full theory are cancelled by its own counterterms).
 In practice, we can avoid dealing with counterterms by  simply using $\overline{MS}$ for any divergence appearing in the full theory calculation (UV and IR), that is, we apply minimal substraction to all the poles in eq.~(\ref{eq:FRfinal}). 
 The above procedure works because we have expanded out any IR scales (fields and external momenta) in both the full and EFT theories. In the presence of explicit light masses in the EFT the preceding discussion is still correct as long as we expand in this mass parameter as well (in both the full and EFT contributions) before integrating over the Schwinger parameters.\footnote{The above discussion is closely related to the so-called "method of regions" \cite{Beneke:1997zp,Smirnov:2002pj,Jantzen:2011nz}. The latter has been applied to read off the one-loop matching contribution from the full theory calculation in the functional formalism \cite{FuentesMartin:2016uol}.}

This concludes the proof of the Feynman rules.
At this point a few comments are in order
\begin{itemize}
\item
The object $\Gamma$ is gauge invariant at each of the vertices, and hence the formula (\ref{eq:FRfinal}) is manifestly gauge invariant. 
Once the partial derivatives are distributed over the different factors of $B_i$ and $C_\ell$ present in $\Gamma$, covariant derivatives will appear, since the product rule is valid for covariant derivatives. 
\item
The new external momenta are not the momenta of single background fields, but rather of a product of many of such fields. The momentum conservation however is the usual one that can be read off of the diagram, that is, an external momentum $p$ entering at a vertex has to be equal to the sum of the momenta exiting through the attached propagators.
\item
The momentum integrations are simple Gaussians, and there are no divergences (neither IR nor UV). However, divergences will appear once the $t_i$ are integrated over. UV divergences appear in the $t_i\to 0$ region, while IR divergences in the $t_i\to\infty$ region. However, if  the momentum integrations are performed in $d$ dimensions, the $t_i$ integrations will be automatically regulated. 
If the Schwinger parameters are integrated before the momenta, the UV and IR divergences appear as usual in the loop integrations.
\item
It is possible to evaluate the Gaussian momentum integral as well as the resulting integrals over the Schwinger parameters without any Wick rotations. However, it is convenient to perform the usual Wick rotation of the momenta together with the following  rotation of the Schwinger parameters
\be
t_i=-i\tau_i\,,
\label{eq:proptimerot}
\ee
with real $\tau_i>0$.
This is allowed as it precisely leaves a real Gaussian momentum integration as well as an integral over Schwinger parameters that is exponentially suppressed for massive propagators $\sim e^{-m_i^2 \tau_i}$.
\item
It is possible to perform the integration over the Schwinger parameters first as in eq.~(\ref{eq:prophk}), in this case the expansion in the external momenta should be done before the loop integrations.
However, we find it more practical to do the $t_i$ integrations at the end, for two reasons.
Firstly, the Schwinger parameters serve as a tool to efficiently combine propagators (similar to Feynman parameters\footnote{Schwinger parameters $t_i$ and Feynman parameters $x_i$ are in fact related by $x_i=t_i/\sum_i t_i$.}) and hence facilitate the treatment of external momenta. Secondly, delaying the expansion of the $B$ factors in $t$ and of the loop amplitude in the externa momenta to the end of the calculation means that the loop momentum integrations are Gaussians that can even be evaluated in closed form in full generality (see app.~\ref{sec:gaussian}).

\end{itemize}

\section{Examples}
\label{sec:examples}

\subsection{A one-loop example}

\begin{figure}
\begin{center}
\includegraphics[width=0.25\linewidth]{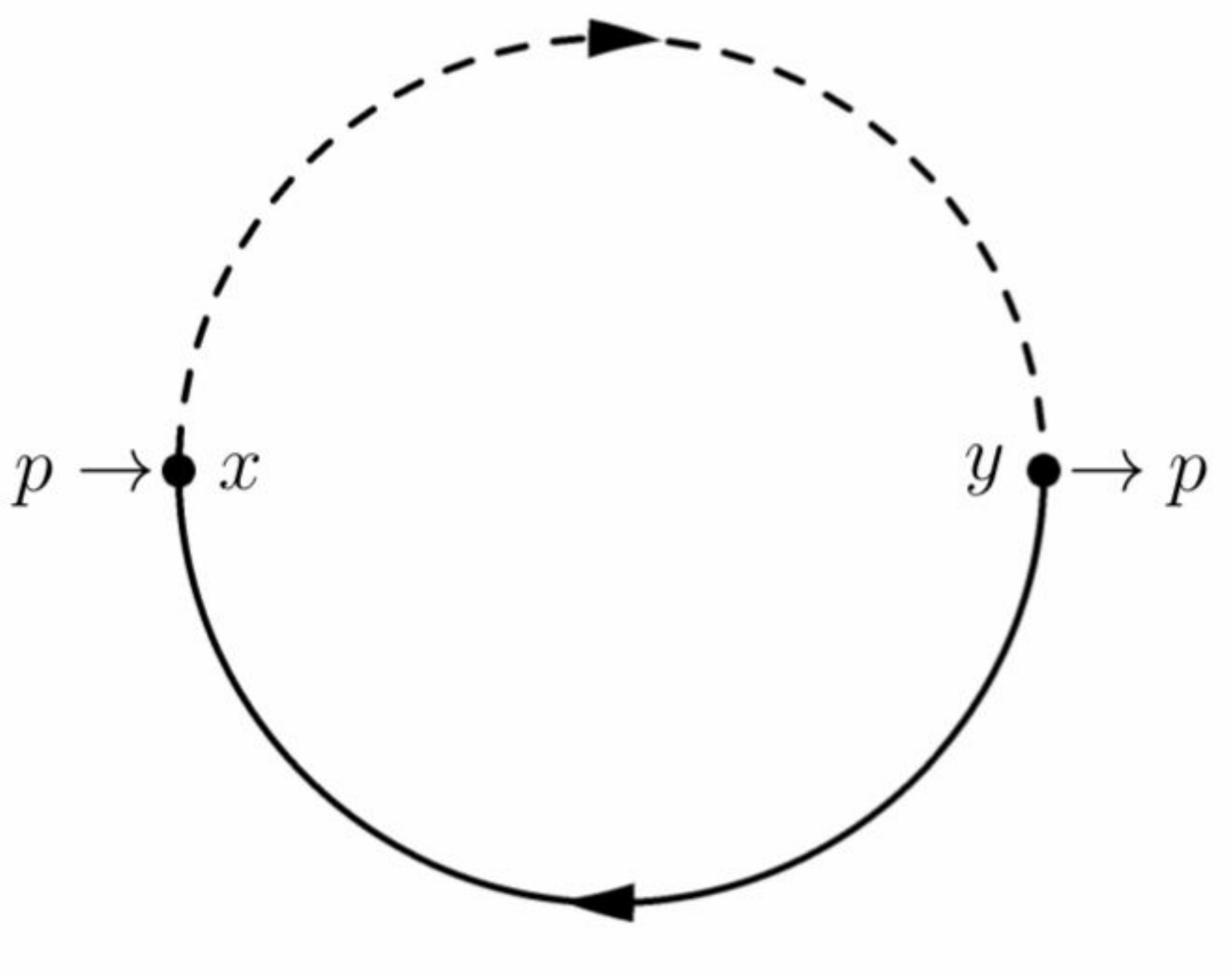}
\end{center}
\caption{A one-loop example.}
\label{fig:oneloop}
\end{figure}

In this section, we integrate out (at one loop) a massive fermion that couples to a light scalar and a light fermion. Let us also focus specifically on the part of $\mathcal L_{\rm eff}$ containing two-fermion operators with additional covariant derivatives and field strengths but no scalar fields. In this case, the only relevant interaction at one-loop is
\be
\mathcal L_{\rm int}=\bar \Psi C \Phi  + h.c.\,,
\ee
where $\Psi$ is the fluctuation of the heavy fermion, $\Phi$ the one of the light scalar, and $C$ is a fermionic field-dependent coupling.
As an example consider a heavy triplet fermion $\Psi^a$ coupled to the SM Higgs $\Phi^i$ and the top quark doublet $q_L^j$, in which case  $C^a_{\ i}(x)=y(\sigma^a)_{ij} q_L^j(x)$ where $y$ is a real coupling. Applying the Feynman rules 
to the diagram in fig.~\ref{fig:oneloop} gives
\bea
i\mathcal L_{\rm eff}&=&\int dt\, ds\,\biggl[\biggr.
\int \frac{d^d k}{(2\pi)^d}\,  e^{i(k^2-m^2) t+i(k+i\partial_x)^2 s}
\nn\\
&& \tr B^s(s;y,x)i\bar C(x)(\sl k+m+i\sl D_x) 
B^f(t;x,y) iC(y) 
 \biggl.\biggr]_{y=x} \,,
\eea
%
where the trace is over the gauge indices, and we have defined the shorthands $B^s\equiv B(X=0)$, $B^f\equiv B(X_f)$. 
Performing the Gaussian momentum integration yields 
\be
\begin{split}
\mathcal L_{\rm eff}=
&\frac{1}{(4\pi)^{\frac{d}{2}}}\int d \s d \t\ \frac{e^{- m^2 \t}}{(\s+\t)^{\frac{d}{2}}}
\\
\biggl[\biggr.
&-i\frac{\s}{\s+\t}\,\exp\left(-\frac{ \s \t}{ \s+ \t}\partial_x^2\right)\, \partial^\mu_x\tr B^s(-i\s;y,x)\bar C(x)\gamma_\mu B^f(-i\t;x,y) C(y) 
\\
&+\exp\left(-\frac{ \s \t}{ \s+ \t}\partial_x^2\right)\, \tr B^s(-i\s;y,x)\bar C(x)(m+ i\sl D_x) B^f(-i\t;x,y) C(y) \biggl.\biggr]_{y=x}\,,
\end{split}
\label{eq:exgeneral}
\ee
We have defined $\t \equiv i t$ and $\s\equiv is$, see discussion around eq.~(\ref{eq:proptimerot}). 
Eq.~(\ref{eq:exgeneral}) is the exact one-loop result to all orders in the operator dimension.
To extract individual operators of a desired dimension, we expand the exponentials containing the partial derivative, as well as $B^f$ and $B^s$ according to eq.~(\ref{eq:hkexp}).
In this particular example, the term proportional to $m$ will yield odd-dimensional operators (zero  for chiral $C$), while the other  terms will yield even dimensional ones. For a nontrivial example, consider the dimension-six operators:
\be
\begin{split}
\mathcal L&^{D=6}_{\rm eff}=
(4\pi )^{-\frac{d}{2}}\int d \s d \t\ \frac{e^{- m^2 \t}}{(\s+\t)^{\frac{d}{2}}}\\
\biggl[\biggr.
&-\frac{\s\t}{\s+\t}\, \partial^\mu_x\tr b_0^s(y,x)\bar C(x)\gamma_\mu b_2^f(x,y) C(y) 
-\frac{\s^2}{\s+\t}\, \partial^\mu_x\tr b_2^s(y,x)\bar C(x)\gamma_\mu b_0^f(x,y) C(y) 
\\
&+i\frac{\s^2\t}{(\s+\t)^2}\,\partial_x^2\partial^\mu_x\tr b^s_0(y,x)\bar C(x)\gamma_\mu b_0^f(x,y) C(y) 
+\t\, \tr b_0^s(y,x)\bar C(x) \gamma^\mu  b^f_{2;\mu}(x,y) C(y) \\
&+\s\, \tr b_2^s(y,x)\bar C(x) \gamma^\mu  b^f_{0;\mu}(x,y) C(y) 
-i\frac{ \s \t}{ \s+ \t}\partial_x^2 \tr b_0^s(y,x)\bar C(x)\gamma^\mu b_{0;\mu}^f(x,y) C(y) \biggl.\biggr]_{y=x}\,,
\end{split}
\ee
where the semicolon denotes covariant differentiation, see eq.~(\ref{eq:covsemi}). Making use of the observation 
pointed out after eq.~(\ref{eq:sympartial}), many terms will be zero, the nonzero ones are:
\be
\begin{split}
\mathcal L&^{D=6}_{\rm eff}=
\frac{1}{16\pi^2}\frac{1}{m^2}
\biggl[\biggr.
-\frac{1}{2} \tr \left(\bar C_{;\mu}\gamma^\mu \cl{b_2^f} C +\bar C\gamma^\mu \cl{b_{2;\mu}^f} C\right)
+\left(\frac{3}{2}-\log\frac{m^2}{\mu^2}\right) \tr \cl{b_{2;;\mu}^{s\phantom f}}\bar C\gamma^\mu  C
\\
&+\frac{i}{3}\tr \bar C_{;(\mu\nu\rho)}\gamma^{\mu}g^{\nu\rho}  C
+ \tr \bar C \gamma^\mu  \cl{b^f_{2;\mu}} C
-\frac{i}{2}g^{\nu\rho} \tr \left(2\bar C_{;\nu}\gamma^\mu \cl{b_{0;\mu\rho}^f} C +
	\bar C\gamma^\mu \cl{b_{0;\mu\nu\rho}^f} C\right) \biggl.\biggr]\,,
\end{split}
\ee
where we already performed the integrals over $\t$ and $\s$ given in App.~\ref{sec:int1}.
The square brackets denote the LHKC's, i.e.~$\cl{b}(x)\equiv b(x,x)$. Notice that at this point, all the $\cl{b}'s$ and $C's$ are evaluated at the same point $x$.
%
%
%
Using the results from app.~\ref{sec:b2n}, and making some simplifications, we arrive at
\begin{multline}
\mathcal L^{D=6}_{\rm eff}=
\frac{1}{16\pi^2}\frac{1}{m^2}
\biggl[\biggr.
\frac{1}{8}\, \tr \bar C_{;\mu}\gamma^\mu (\sl F^f) C  -\frac{1}{8} \tr C\sl F^f\gamma^\mu C_{;\mu}
+\frac{i}{3}\,\tr \bar C_{;(\mu\nu\rho)}\gamma^{\mu}g^{\nu\rho}  C\\
+\left(\frac{1}{4}-\frac{1}{6}\log\frac{m^2}{\mu^2}\right)\, \tr F^s_{\mu\nu;\nu}\bar C\gamma^\mu  C
+ \frac{1}{4}\tr \bar C \gamma^\mu  F^f_{\mu\nu;\nu} C
\biggl.\biggr]\,.
\end{multline}
The calculation of these dimension-six operators in non-covariant perturbation theory would require the evaluation of 14 diagrams (not counting permutations of external legs), including some tedious Dirac algebra, plus the reconstruction of the covariant operators in terms of the noncovariant result.

\subsection{A two-loop example}

\begin{figure}
\begin{center}
\includegraphics[width=0.25\linewidth]{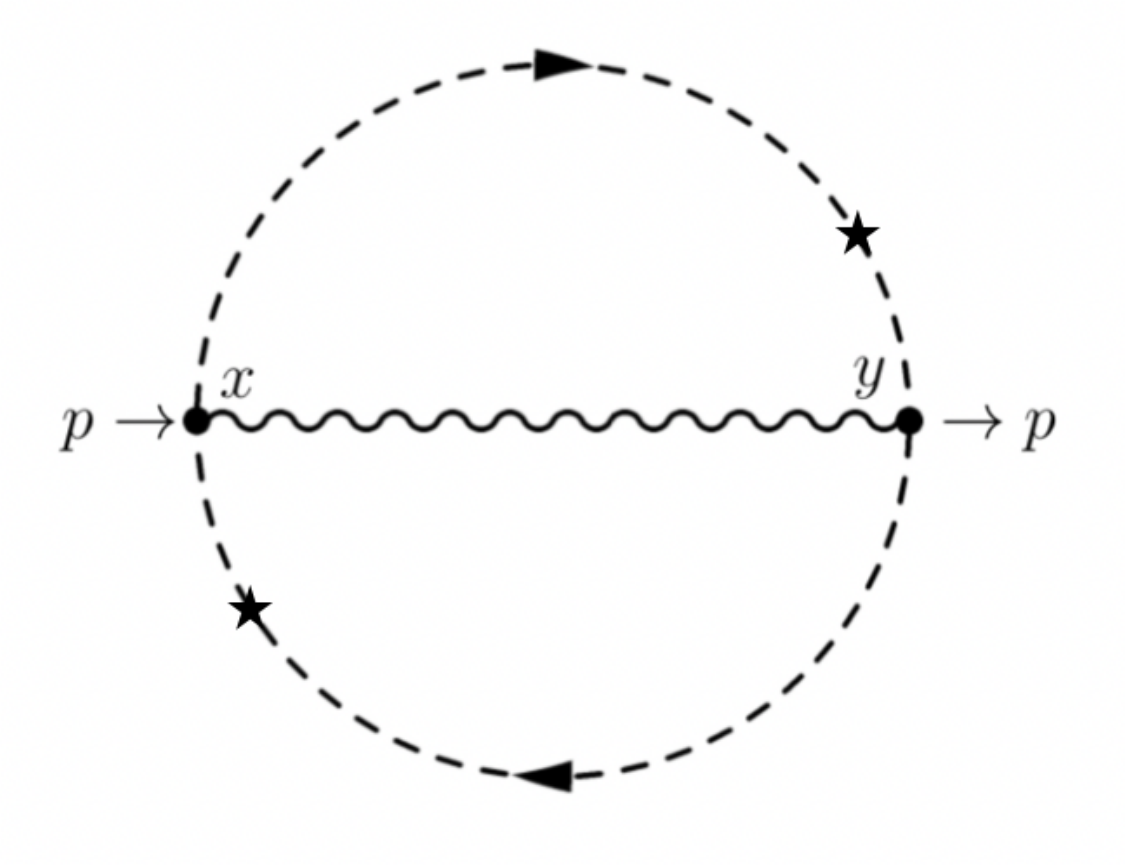}
\includegraphics[width=0.25\linewidth]{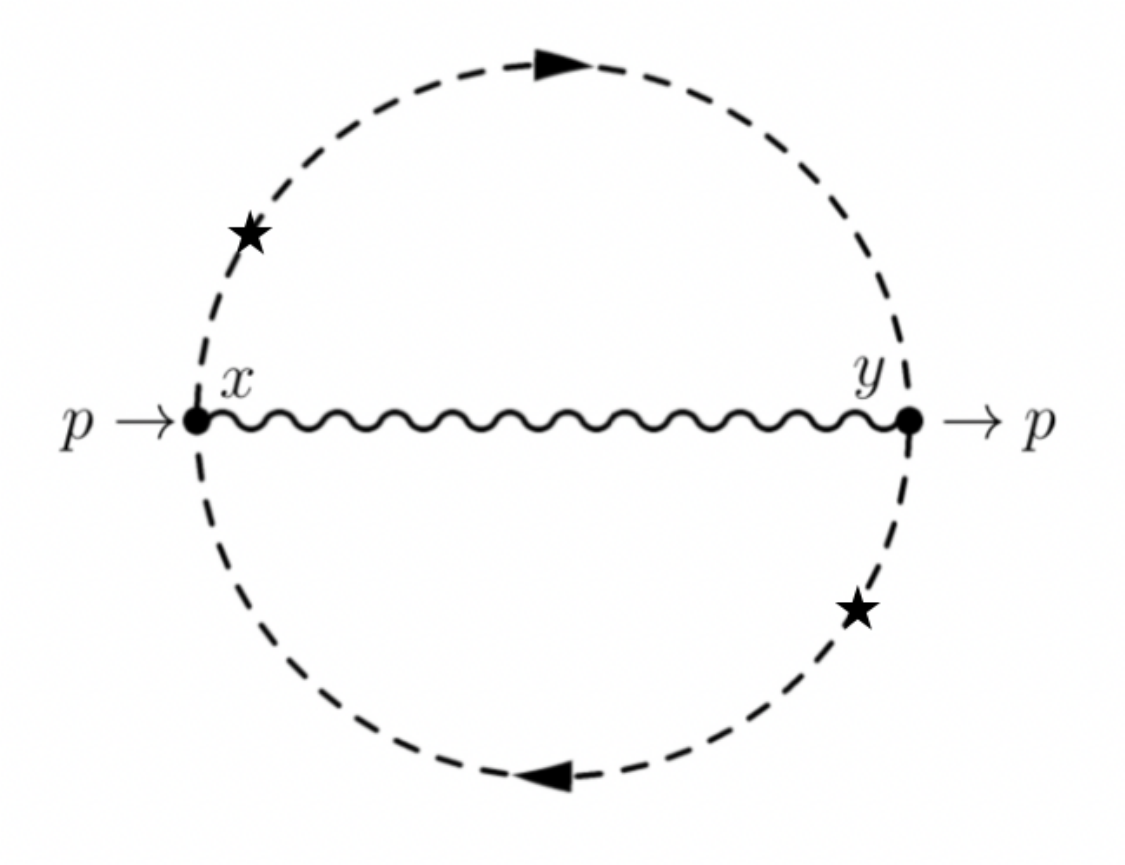}
\includegraphics[width=0.25\linewidth]{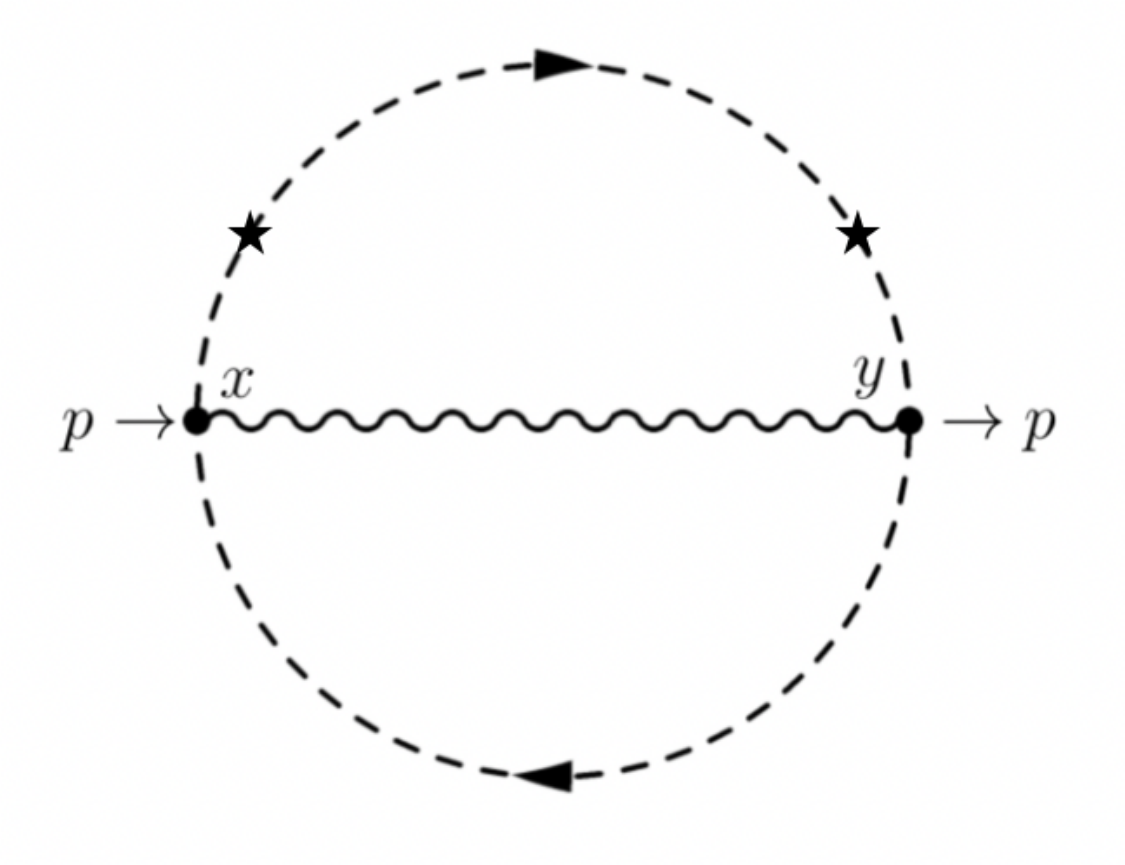}
\ \
\includegraphics[width=0.09\linewidth]{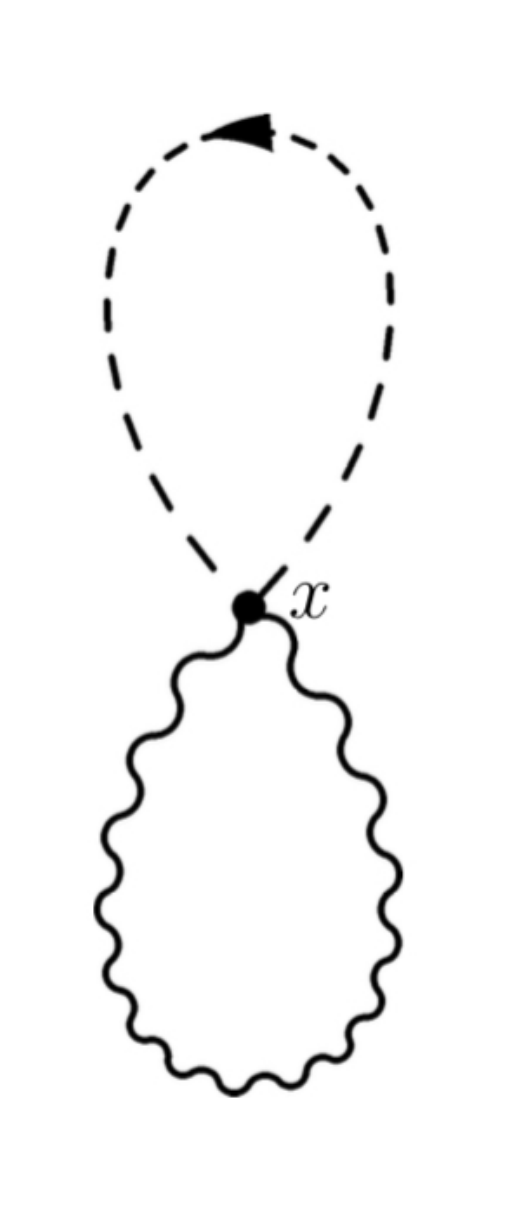}
\end{center}
\caption{A two-loop example. The first, second and fourth diagrams carry a symmetry factor of $\frac{1}{2}$. The stars denote a derivative on a field.}
\label{fig:twoloop}
\end{figure}

In this section we would like to apply our formalism to integrate out a massive scalar coupled to gauge theory at two loops. 
As should be clear by now, our formalism contains roughly  three main calculational steps: the Gaussian momentum integration, the reduction to local operators, and the integral over the Schwinger parameters.
We will see that the first two steps are virtually identical at all loops, in particular, the momentum integration can be given in closed form, see eq.~(\ref{eq:I2}).
The integrals over the Schwinger parameters become more involved at higher loops, in the example provided below they can however still be computed analytically.

 Assuming that the heavy scalar does not acquire a vacuum expectation value, we can set the scalar background field to zero, that is, all the bilinear vertices in tab.~\ref{tab:bil} vanish in any gauge.
The two-loop diagrams with at least one heavy propagator are the three sunset graphs and the figure-eight graph  shown in fig.~\ref{fig:twoloop}.

The figure-eight diagram gives
\be
i\mathcal L^{\rm fig-8}_{\rm eff}=\frac{1}{2}
\int ds dt
\int\frac{d^dk}{(2\pi)^d}\frac{d^dk'}{(2\pi)^d}\, e^{ik^2s+ik'^2t-im^2s} 
\tr\left\{(ig^2)\{\gen^a,\gen _b\}B(s;x,x)\right\}B^{\mu b}_{\ \mu a}(t;x,x)
\ee
where the symmetry factor of $\frac{1}{2}$ results from the tadpole loop of the gauge boson.
Performing the momentum integration yields
\be
\mathcal L_{\rm eff}^{\rm fig-8}=\frac{g^2}{2(4\pi)^d}
\int d\s 
\, e^{-m^2\s} \s^{-\frac{d}{2}}
\tr\left\{\{\gen^a,\gen _b\}B(-i\s;x,x)\right\}\int d\t\,\tau^{-\frac{d}{2}} B^{\mu b}_{\ \mu a}(-i\t;x,x)
\ee
Being a scaleless integral the $\tau$ integration gives zero in dimensional regularization.

Let us then evaluate the three sunset graphs. 
These diagrams also illustrate how derivative interactions are treated in our formalism. We mark a covariant derivative of a field by a star. The triple vertex connects to one scalar and one scalar with covariant derivative and hence each vertex carries one star, and the three diagrams represent the different ways of placing the stars  in the diagram. The first two diagrams carry a symmetry factor of $\frac{1}{2}$ because the two vertices are equivalent, whereas the third diagram has no symmetry factor (the two vertices are not even the same, but rather complex conjugates).\footnote{We further elaborate on symmetry factors in app.~\ref{sec:fluc}.}

Their contribution to the effective Lagrangian is
\be
\begin{split}
i\mathcal L_{\rm eff}=\int & drdsdt
\biggl[\biggr.
\int\frac{d^dk}{(2\pi)^d}\frac{d^dk'}{(2\pi)^d}\, e^{ik^2r+ik'^2s+i(k-k'+i\partial_x)^2t-im^2(s+r)}
\\
\biggl(\biggr.
 \frac{1}{2}&\tr \left\{(g \gen^a)(-ik_\mu +D^x_\mu)B(r;x,y)(g\gen^b)(-ik'_\nu+D_\nu^y)B(s;y,x)\right\} 
\\
+ \frac{1}{2}&\tr \left\{(-g \gen^a)(ik_\nu +D^y_\nu)B(r;x,y)(-g\gen^b)(ik'_\mu+D_\mu^x)B(s;y,x)\right\} 
\\
+& \tr \left\{(-g \gen^a)B(r;x,y)(g\gen^b)(-ik'_\nu+D_\nu^y)(ik'_\mu +D^x_\mu)B(s;y,x)\right\} 
\biggl.\biggr)B^{\mu\nu}_{ab}(t;x,y)\biggl.\biggr]_{y=x}
\end{split}
\ee
The three traces correspond to the three diagrams. 
Notice how extra factors of $-ik_\mu+D_\mu$ etc.~appear due to the propagators of derivatives of fields (denoted by stars in the diagrams), these derivatives are meant only to act on the $B$ factor immediately to the right (in contrast  the partial derivative in the exponent which acts on all three $B$ factors).
The Gaussian momentum integration is straightforward, 
see the master formula Eq.~(\ref{eq:I2}).
Focusing for instance on the first diagram, one gets
\be
\begin{split}
\mathcal L^{(1)}_{\rm eff}=-&\frac{g^2}{2} \frac{1}{(4\pi)^{d}} \int d\r d\s d\t\, \frac{e^{-m^2(\s+\r)}}{\omega^{\frac{d}{2}}}
\\
\biggl[\biggr.
&e^{-\frac{\r\s\t}{\omega}\partial_x^2}
	\left(\frac{\tau}{2\omega}g_{\mu\nu}-\frac{\r\s\t^2}{\omega^2}\partial_\mu^x\partial_\nu^x
	\right) \tr \left\{\gen^aB(-i\r;x,y)\gen^bB(-i\s;y,x)\right\} B^{\mu\nu}_{ab}(-i\t;x,y)\\
+&e^{-\frac{\r\s\t}{\omega}\partial_x^2} \left(\frac{\r\t}{\omega}\partial^x_\nu\right) \tr \left\{\gen^a D^x_\mu B(-i\r;x,y)\gen^bB(-i\s;y,x)\right\} B^{\mu\nu}_{ab}(-i\t;x,y)\\
+&e^{-\frac{\r\s\t}{\omega}\partial_x^2} \left(-\frac{\s\t}{\omega}\partial^x_\mu \right)\tr \left\{\gen^aB(-i\r;x,y)\gen^bD_\nu^yB(-i\s;y,x)\right\} B^{\mu\nu}_{ab}(-i\t;x,y)\\
+&e^{-\frac{\r\s\t}{\omega}\partial_x^2} \tr \left\{\gen^a D^x_\mu B(-i\r;x,y)\gen^b D_\nu^y B(-i\s;y,x)\right\} B^{\mu\nu}_{ab}(-i\t;x,y)
\biggl.\biggr]_{y=x}
\end{split}
\ee
where $\omega\equiv\t\s+\s\r+\r\t$.
Again this is the all-order result in the operator dimension. The other two diagrams are completely analogous.
The next step would be to expand this result up to a desired order in the operator dimension, that is, perform the expansion of the exponentials containing the derivatives as well as the expansion eq.~(\ref{eq:hkexp}).
There is nothing new to say here, this expansions work exactly as in the one-loop example.
The resulting intergrals over the Schwinger parameters can still be performed analytically, see app.~\ref{sec:int2}.

\section{Conclusions}

We have presented new covariant Feynman rules that allow for a simple and efficient calculation of the local effective Lagrangian from integrating out heavy particles.
The formalism is universal with no restrictions to the type of interactions (including arbitrary derivative interactions), loop order, or type of particles in the loop (e.g.~spin, massive, massless).

Even though our main focus was on the EFT matching contributions, our formalism can equally well be applied elsewhere, for instance to the computation of renormalization group functions of general effective field theories.

A possible generalization is the inclusion of a gravitational background, many results on the gravitational LHKC's exist in the literature, see ref.~\cite{Vassilevich:2003xt} for an overview. 

Our algoithm is amenable to full automatization and we have provided a first small step in that direction by including with this publication a mathematica notebook that computes the LHKC's.

\appendix
\section{Background field method for gauge theories}
\label{sec:fluc}

\newcommand{\A}{{\mathcal A}}

In this appendix we will derive the field dependent couplings for gauge theories. 
We include scalar fields here due to their entanglement in the gauge fixing procedure.
Let us thus consider the Lagrangian
\be
\mathcal L=-\frac{1}{4 g^2}   (\hat F^a_{\mu\nu})^2 +|\hat D_\mu\hat \phi|^2\,.
\label{eq:gaugescalar}
\ee
One decomposes the fields into backgrounds ($\phi$, $A$) and fluctuations ($\phih$, $\Ah$)\footnote{We work with the convention $D_\mu=\partial_\mu-iA^a_\mu \gen^a$ such that
$
[D_\mu,D_\nu]=
-iF^a_{\mu\nu}\gen^a
$. Furthermore the adjoint generators are $(\gen_{adj}^a)_{bc}=-if_{abc}$
}
\bea
\hat \phi&=& \phi+\phih\,,\nn\\
\hat A^a_\mu&=& A^a_\mu+g\Ah^a_\mu\,,\\
\hat F^a_{\mu\nu}&=& F^a_{\mu\nu}+gD_\mu \Ah^a_\nu-gD_\nu \Ah^a_\mu+g^2f^{abc}\Ah^b_\mu\Ah^c_\nu\nn\,.
\eea
To eq.~(\ref{eq:gaugescalar}) we add the following gauge fixing Lagrangian
\be
\mathcal L^{\rm g.f.}=-\frac{1}{2\xi }(D_\mu  \Ah_\mu^a+ \eta\xi [i\phih^\dagger \Y^a+h.c.])^2\,,
\ee
where we defined 
\be
\Y^{i a}\equiv g(\gen^a)^i_{\ j}\phi^j\,,
\ee
 and $\xi$ and $\eta$ are real gauge fixing parameters. 

One finds the following terms quadratic, cubic and quartic in the fluctuations:
\bea
\mathcal L_2&=&-\frac{1}{2} (D_\mu \Ah^a_\nu)^2+\frac{1}{2}\left(1-\xi^{-1}\right) (D^\mu  \Ah^a_\mu)^2
		- f^{abc} F^a_{\mu\nu}\Ah^b_\mu \Ah^c_\nu +\Ah_\mu^a\Ah_\mu^b(\Y^\dagger \Y)^{ab}\nn \\
			&&+\,|D_\mu \phih|^2
			-{\eta^2\xi}\,\phih^\dagger\Y\Y^\dagger\phih
			\label{eq:bil}\\
			&&+\left(\frac{\eta^2\xi}{2} \phih^T\Y^*\Y^\dagger \phih
			+i\Ah_\mu^a\left(\eta+1\right)\phih^\dagger D^\mu\Y^a
			+i\Ah_\mu^a\left(\eta-1\right)D^\mu \phih^\dagger\Y^a+h.c\right)\,,
	\nn	\\
\mathcal L_3&=& -g f^{abc}D_\mu \Ah^a_\nu \Ah^b_\mu \Ah^c_\nu+g\Ah_\mu^a(i\phih^\dagger \gen^a D^\mu \phih+h.c. )\,,
	\label{eq:tri}\\
\mathcal L_4&=&-\frac{g^2}{4}f^{eab}f^{ecd}\Ah^a_\mu\Ah^b_\nu\Ah^c_\mu\Ah^d_\nu
	+g^2\Ah_\mu^a\Ah_\mu^b\phih^\dagger \gen^a\gen^b\phih \,.
	\label{eq:quad}
\eea

For the ghost Lagrangian, one obtains
\be
\mathcal L^{\rm ghost}=-\bar c^a\left(\delta^{ab} D^2
+2\eta\xi\, (\Y^\dagger \Y)^{(ab)}
-g\overleftarrow D_\mu f^{acb}\Ah^\mu_c
+g\eta\xi\, (\phih^\dagger \gen^b \Y^a+h.c.)
\right)c^b\,.
	\label{eq:ghost}
\ee
The gauge choice $\xi=1$ eliminates the  non minimal terms in the gauge propagator.
In addition, most often the choice $\eta=1$ is adopted, in order to remove the bilinear 
term  $\sim\A_\mu D^\mu\Phi$. However, as is clear from the above Lagrangians,  $\eta\neq 0$ comes at a price, as we are generating many new terms in the above Lagrangians.
Since our formalism deals very naturally with derivatives on propagators,  other choices, in particular $\xi=1, \eta=0$, seem at least equally motivated. 
The last line in eq.~(\ref{eq:bil}) are bilinear vertices, they cannot be eliminated completely for any choice fo $\xi$ and $\eta$.
From eq.~(\ref{eq:bil}) and eq.~(\ref{eq:ghost}) one reads off the field dependent masses for scalars, vectors and ghosts:
\bea
(X_s)^i_{\ j}&=&\eta^2\xi G^{ia}G^*_{ja}\,,\label{eq:Xs}\\
(X_v)^{a\,\mu}_{\ b\,\nu}&=& -2f^{a}_{\ bc}F^{c\,\mu}_{\ \  \nu}
+
(G^{ia}G^*_{ib}+G^*_{ia}G^{ib})\delta^\mu_{\ \nu}\,.\\
(X_{g})^{a}_{\ b}&=&
\eta \xi (G^{ia}G^*_{ib}+G^*_{ia}G^{ib})
\eea
From the last line of eq.~(\ref{eq:bil}) as well as eqns.~(\ref{eq:tri}), (\ref{eq:quad}), and (\ref{eq:ghost}) we can read off the Feynman rules for the field dependent vertices, which we summarize in tables \ref{tab:bil}, \ref{tab:tri} and \ref{tab:quad}.

\begin{table}
\begin{center}
\begin{tabular}{|c|c|}
\hline
&\\
\includegraphics*[width=5.1 cm]{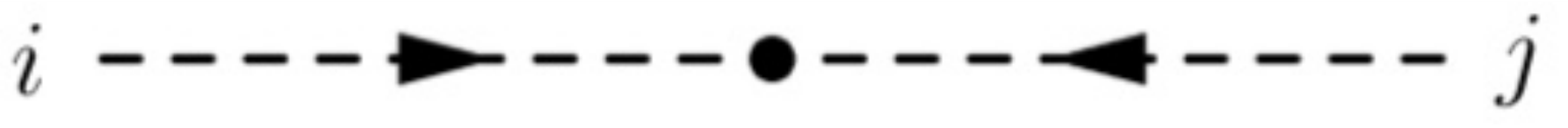}
&\includegraphics*[width=5.1 cm]{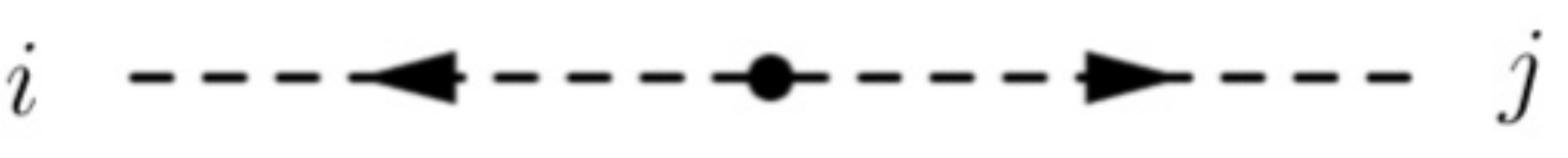}
\\
	 $i\eta^2\xi \, \Y^*_{ia}\Y^*_{ja}$
	& $i\eta^2\xi \, \Y^{ia}\Y^{ja}$\\
&\\
	\hline	
&\\	
\includegraphics*[width=5.4cm]{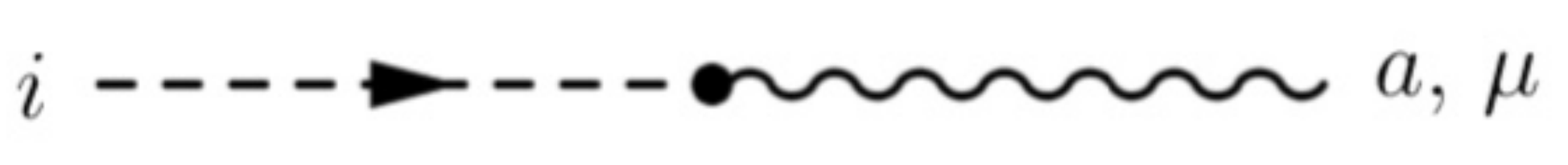}
&\includegraphics*[width=5.4cm]{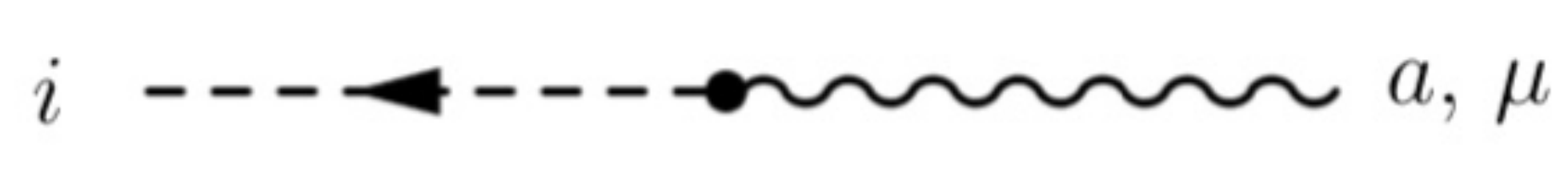}
\\
	 $(\eta+1)D_\mu \Y^*_{ia}$
	& $-(\eta+1)D_\mu \Y^{ia}$\\
&\\
\hline
&\\
\includegraphics*[width=5.4cm]{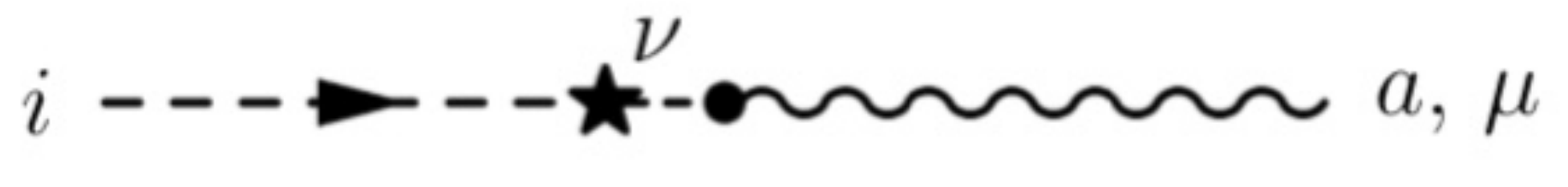}
&\includegraphics*[width=5.4cm]{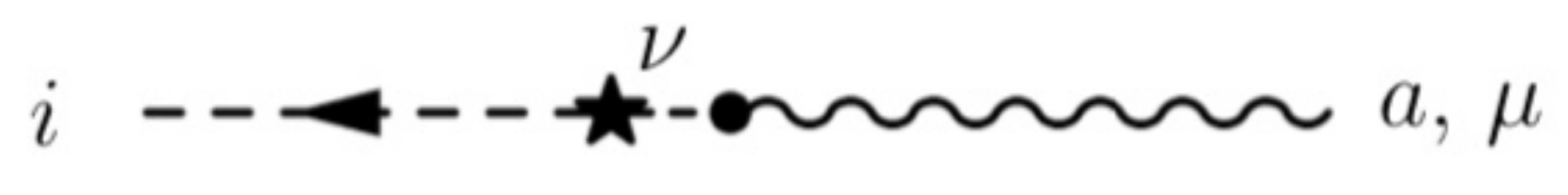}
\\
	 $(\eta-1)g_{\nu\mu}G^*_{ia}$
	& $-(\eta-1)g_{\nu\mu}G^{ia}$\\	
&\\
	\hline
\end{tabular}
\end{center}
\caption{Bilinear vertices for gauge theory coupled to a scalar. Recall that $G^{ia}=g(\gen^a)^i_{\ j}\phi^j$. An asterisk with index $\nu$ indicates that the attached propagator is a derivative one, i.e., $\braket{\Phi_{;\nu} \Phi^\dagger }$ (for ingoing arrow) or $\braket{\Phi \Phi_{;\nu}^\dagger}$ (for outgoing arrow). 
The right and left columns are complex conjugate vertices.}
\label{tab:bil}
\end{table}

\begin{table}[h]
\begin{center}
\begin{tabular}{|c|c|}
\hline
&\\
\includegraphics*[width=4.5cm]{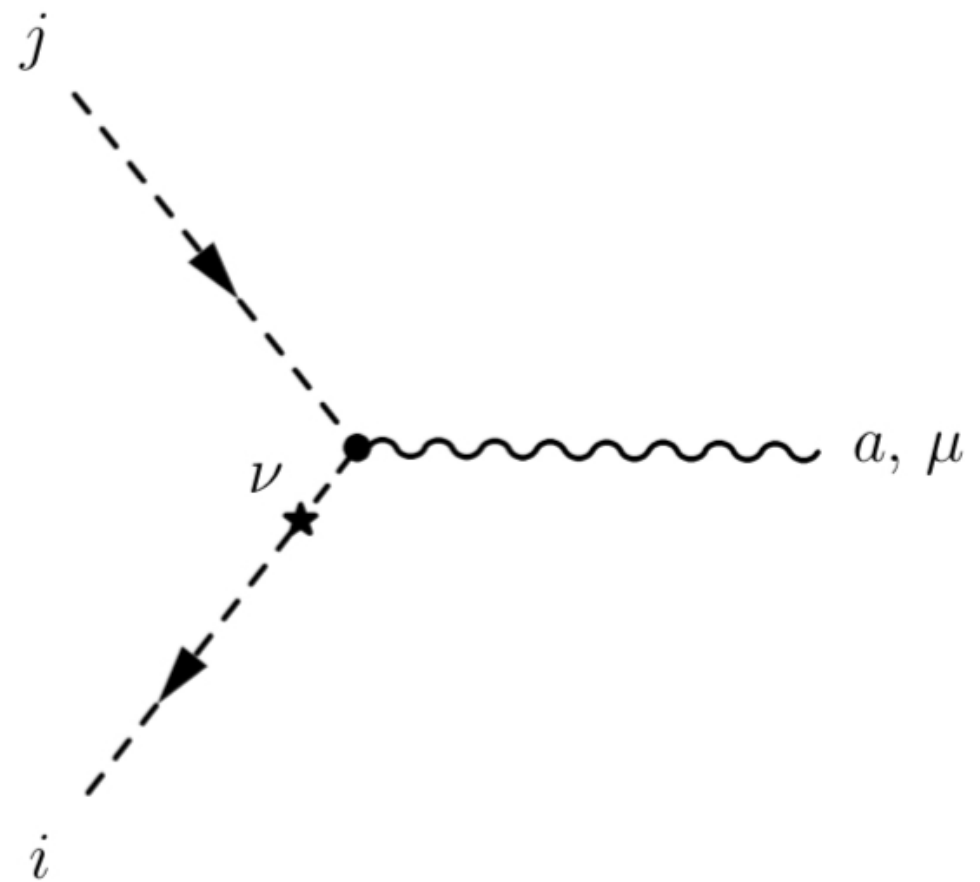}
&\includegraphics*[width=4.5cm]{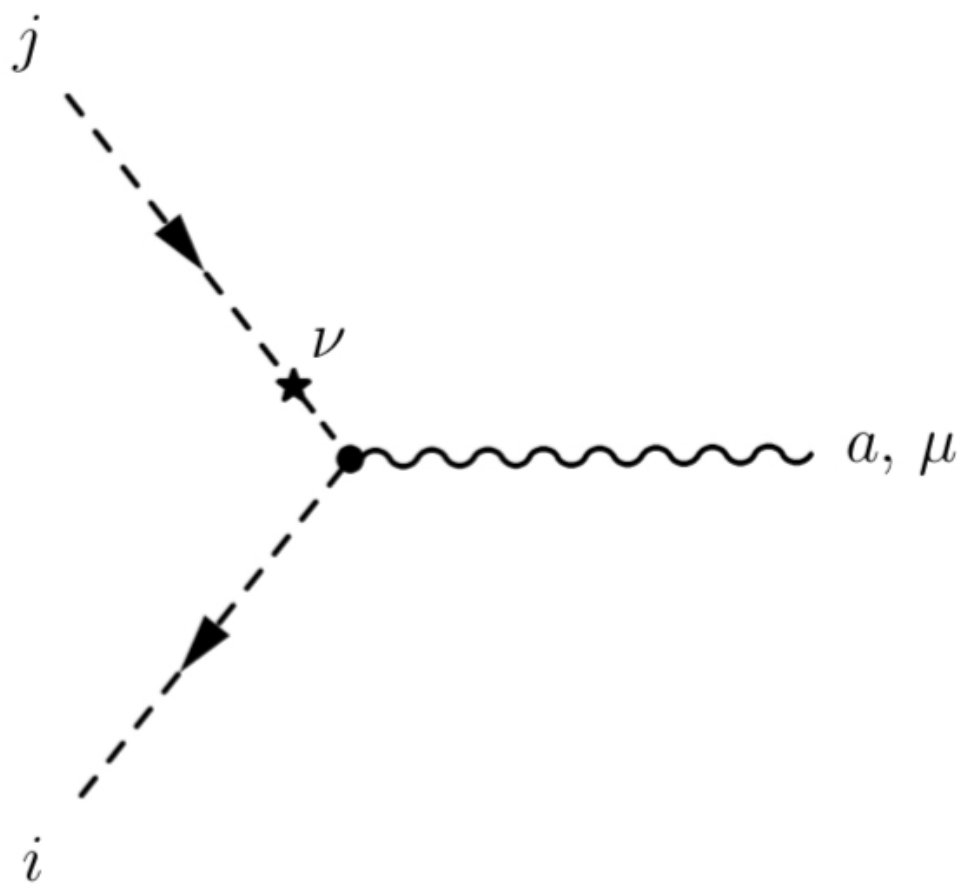}
\\
	 $-g(\gen^a)^i_{\ j}g_{\nu\mu}$
	& $g(\gen^a)^i_{\ j}g_{\nu\mu}$\\
&\\
	\hline	
&\\	
\includegraphics*[width=4.5cm]{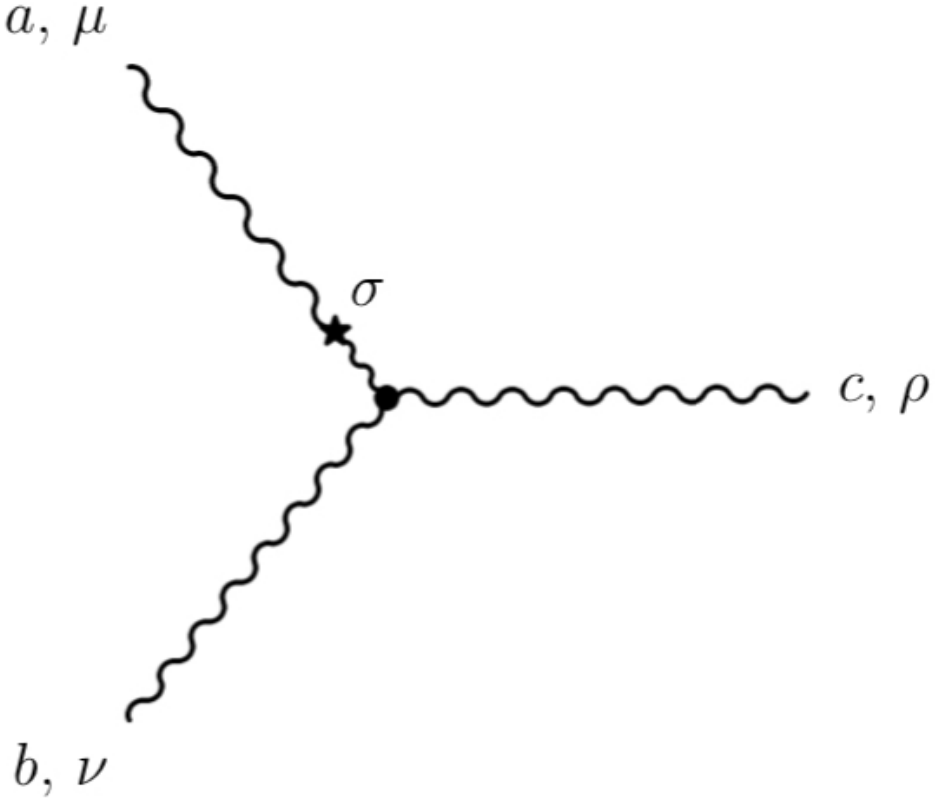}
&\includegraphics*[width=4.5cm]{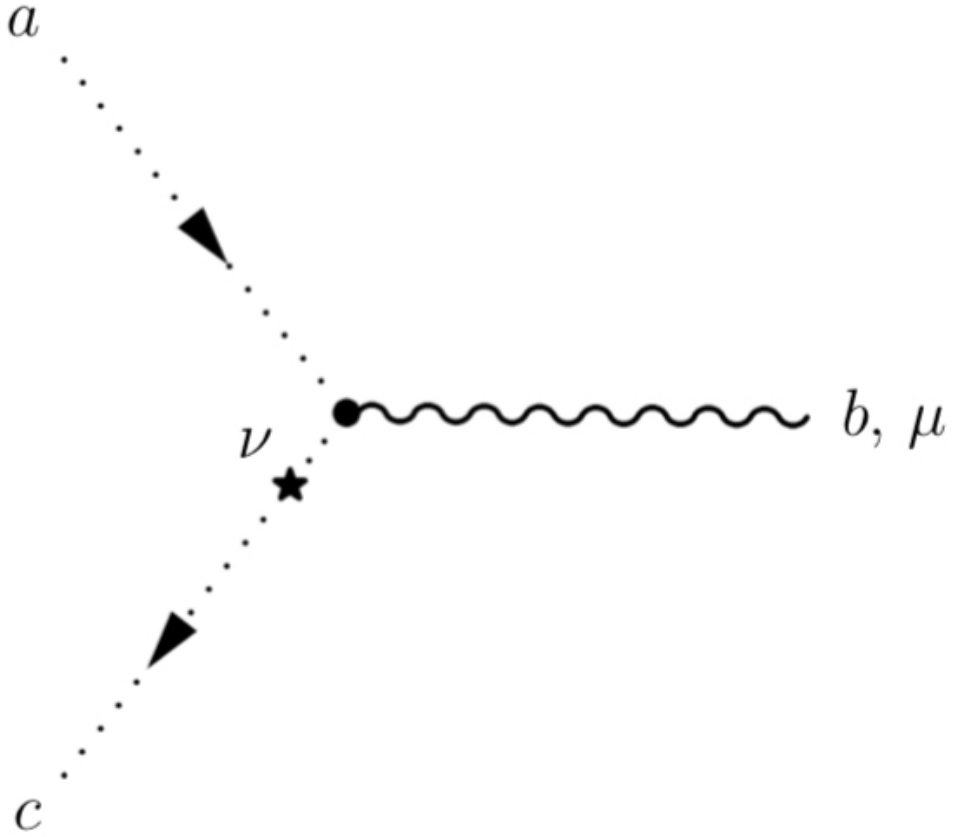}
\\
	 $i g f_{abc}(g_{\mu\nu}g_{\sigma\rho}-g_{\mu\rho}g_{\sigma\nu})$
	& $-igf_{abc}g_{\mu\nu}$\\
&\\
\hline
&\\
\includegraphics*[width=4.5cm]{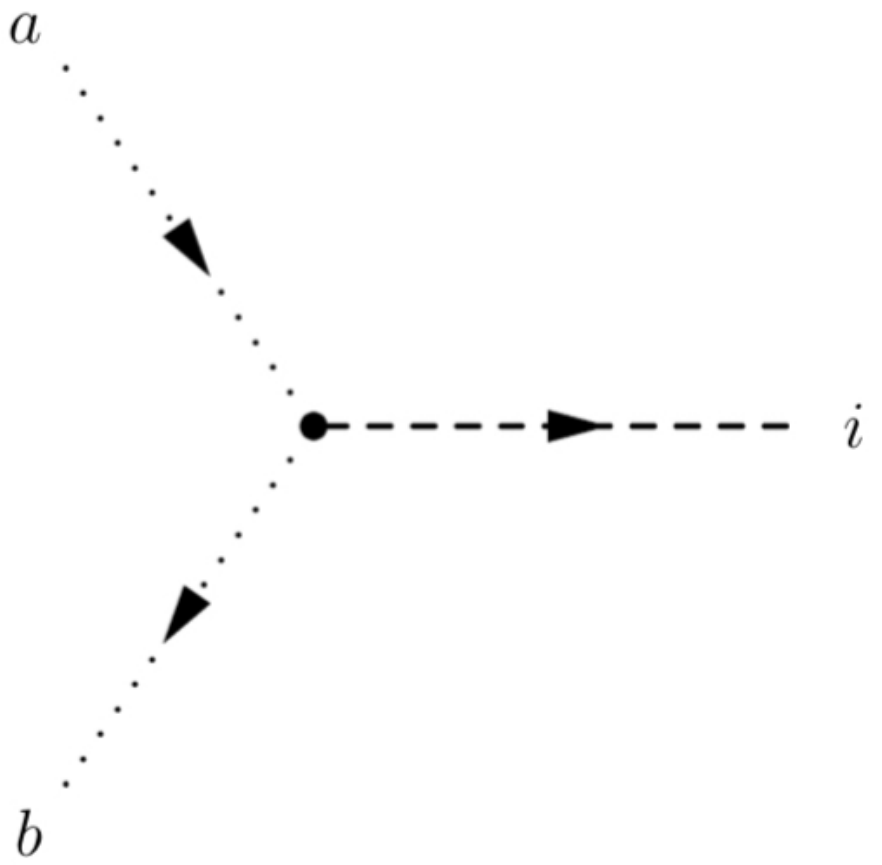}
&\includegraphics*[width=4.5cm]{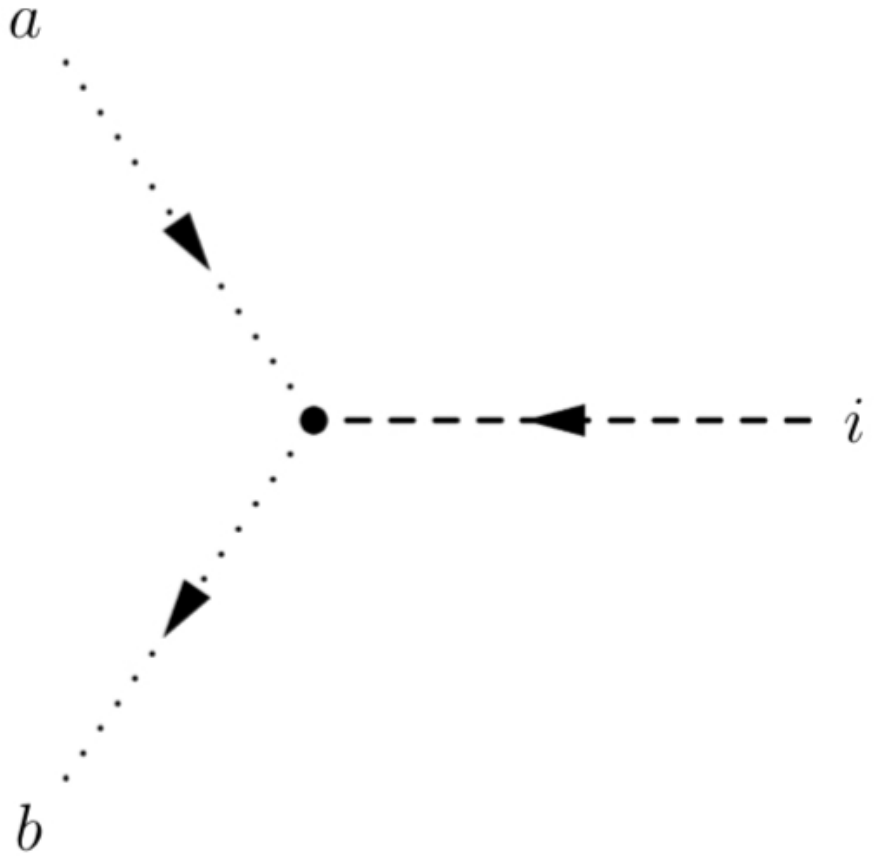}
\\
	 $-ig\eta\xi (\gen^a)^i_{\ j} G^{jb}$
	& $-ig\eta\xi (\gen^a)^i_{\ j} G^*_{ib}$\\	
&\\
	\hline
\end{tabular}
\end{center}
\caption{Trilinear vertices for gauge theory coupled to a scalar. Recall that $G^{ia}=g(\gen^a)^i_{\ j}\phi^j$. An asterisk  indicates that the attached propagator is a derivative one, i.e., $\braket{\Phi_{;\nu} \Phi^\dagger }$ 
(for ingoing arrow), $\braket{\Phi \Phi_{;\nu}^\dagger}$ (for outgoing arrow), $\braket{A_{\mu;\sigma}A_\lambda}$, or $\braket{c \, \bar c_{;\nu}}$.}
\label{tab:tri}
\end{table}

\begin{table}[h]
\begin{center}
\begin{tabular}{|c|c|}
\hline
&\\
\includegraphics[width=4.5cm]{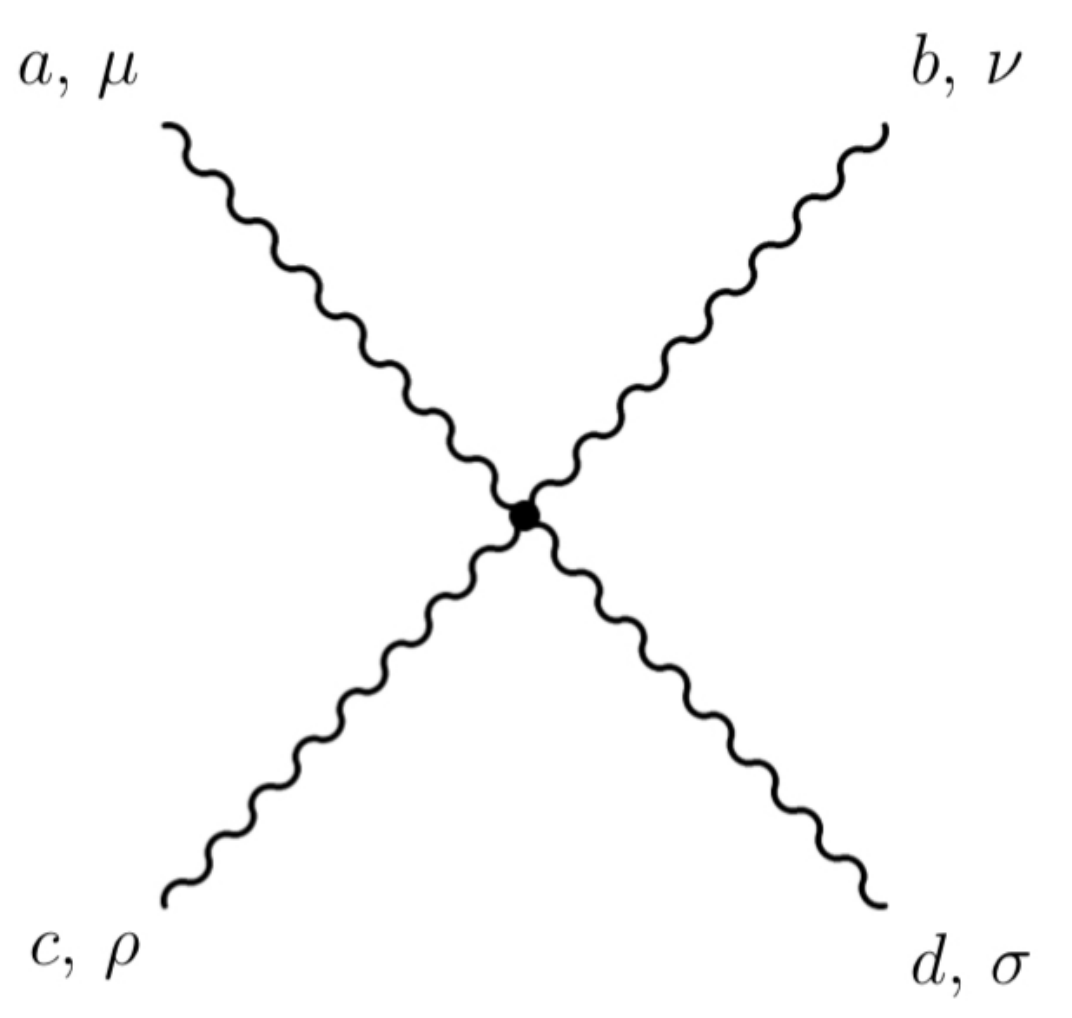}&
\includegraphics[width=4.5cm]{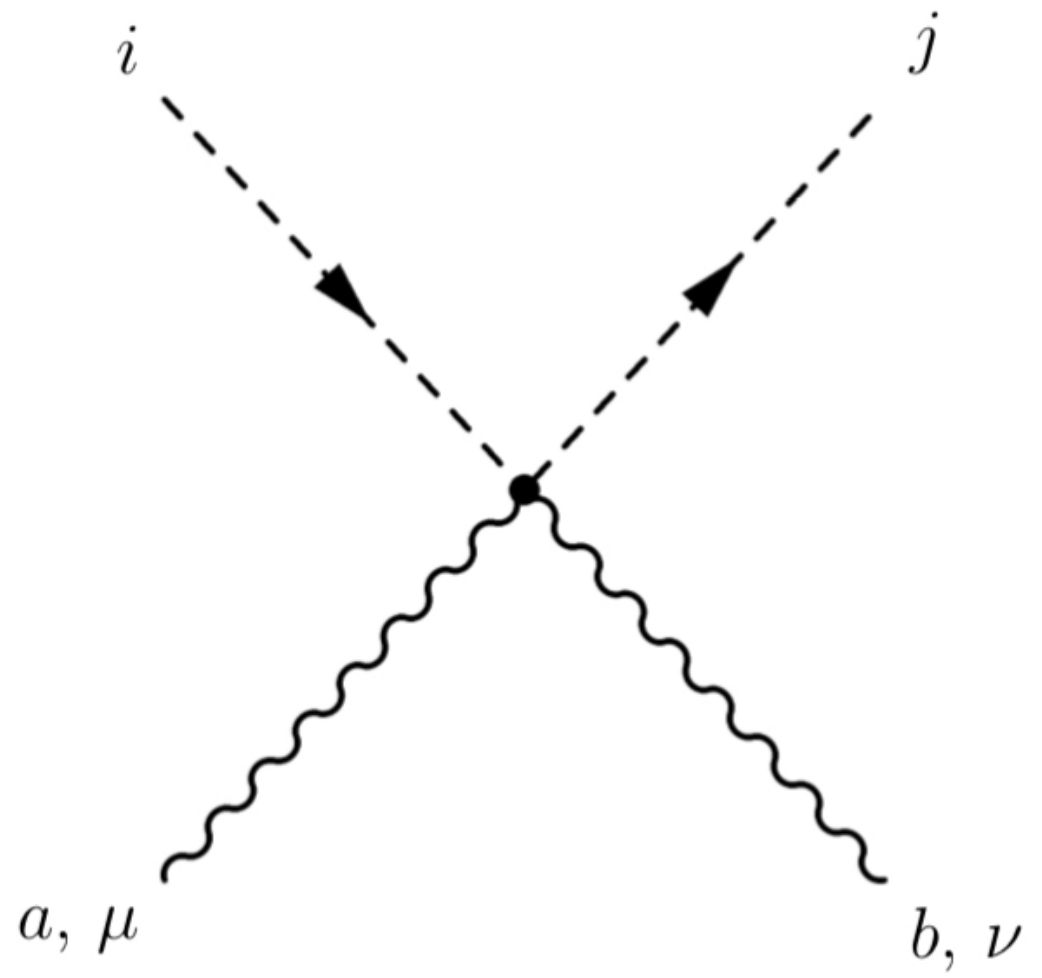}
\\
$
-ig^2 [f^{abe}f^{cde}(g^{\mu\rho} g^{\nu\sigma}-g^{\mu\sigma} g^{\nu\rho})
$
&$i g^2 g_{\mu\nu}\, \{\gen^a,\gen^b\}^i_{\ j}$
\\
$\phantom{-ig^2[}f^{ace}f^{bde}(g^{\mu\nu} g^{\rho\sigma}-g^{\mu\sigma} g^{\nu\rho})$&\\
$\phantom{-ig^2[}f^{ade}f^{bce}(g^{\mu\nu} g^{\rho\sigma}-g^{\mu\rho} g^{\nu\sigma})]$&\\
&\\
\hline
\end{tabular}
\end{center}
\caption{Quartic vertices for gauge theory coupled to a scalar. There is no difference to the usual non-covariant approach, and they are only included here for completeness.}
\label{tab:quad}
\end{table}

\begin{figure}[h]
\begin{center}
\begin{tabular}{cccc}
\includegraphics[width=3.2cm]{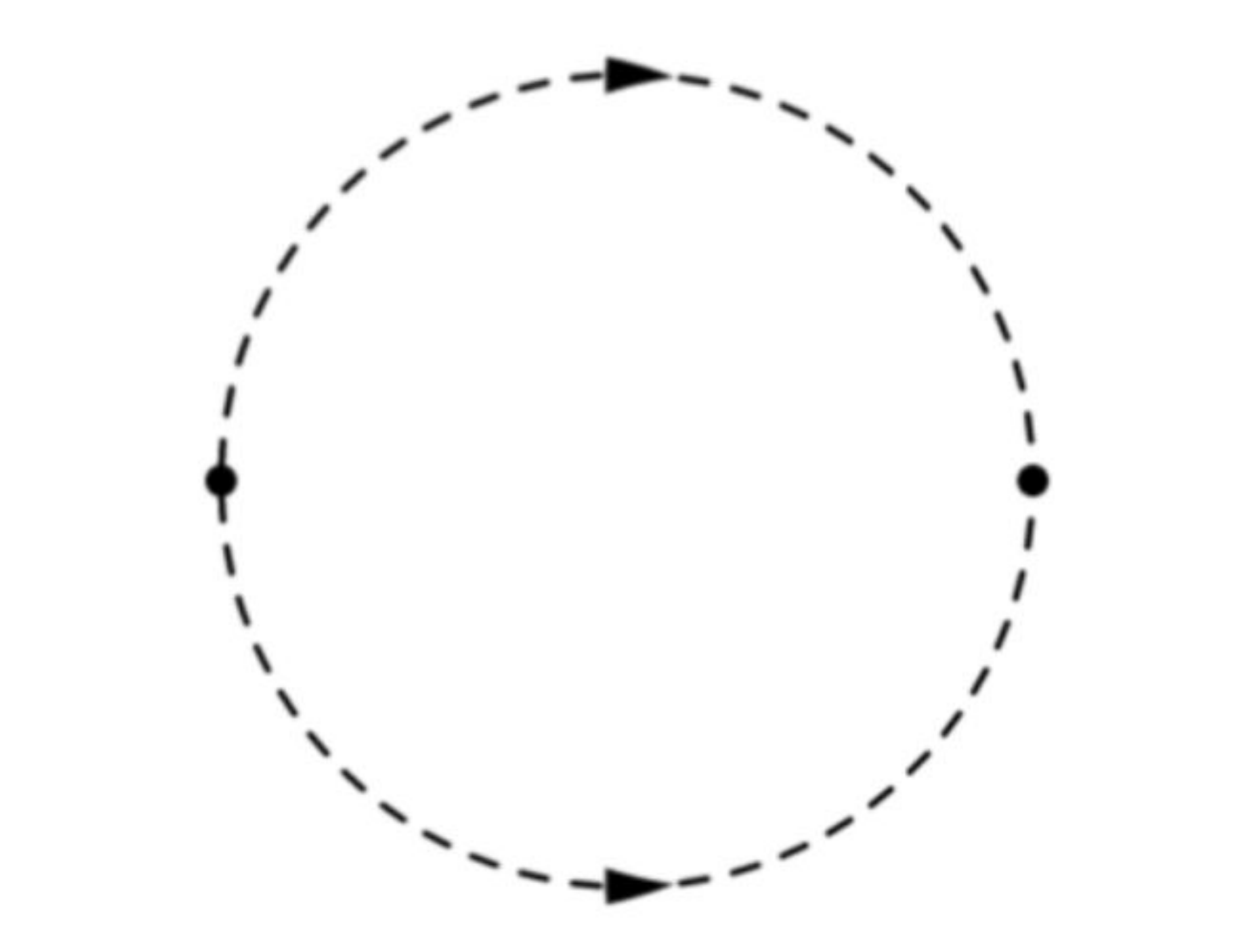}
&\includegraphics[width=3.2cm]{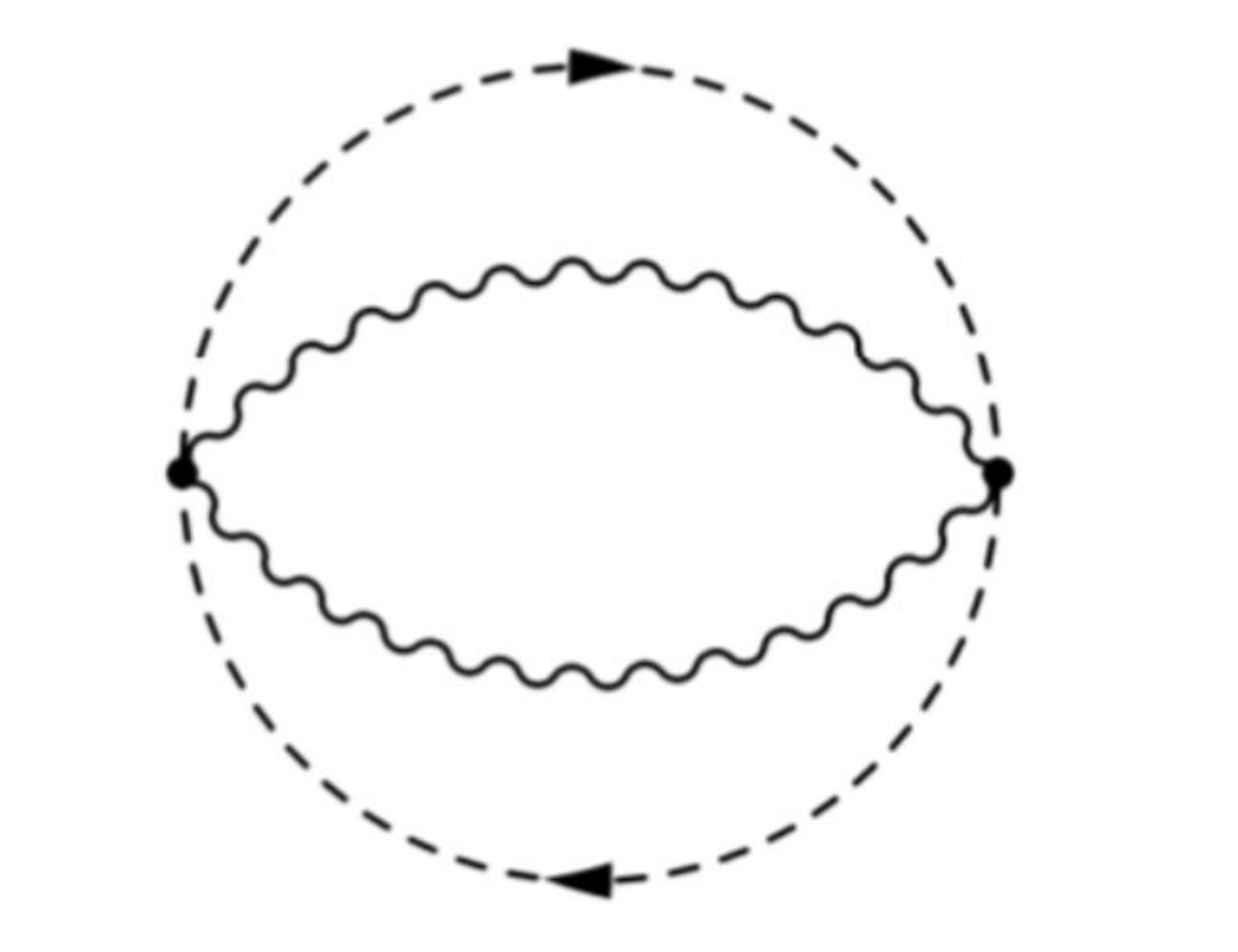}
&\includegraphics[width=3.2cm]{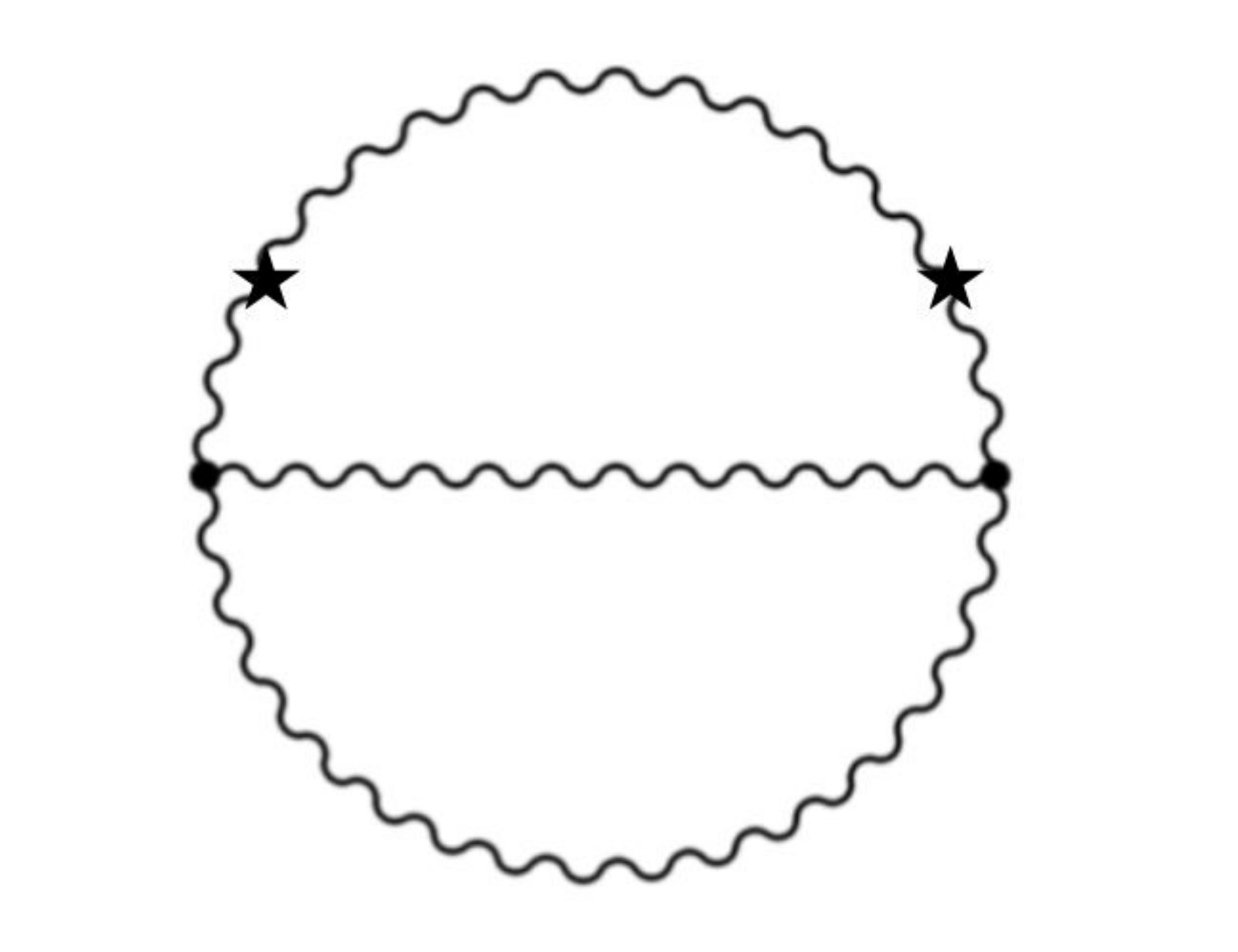}
&\includegraphics[width=3.2cm]{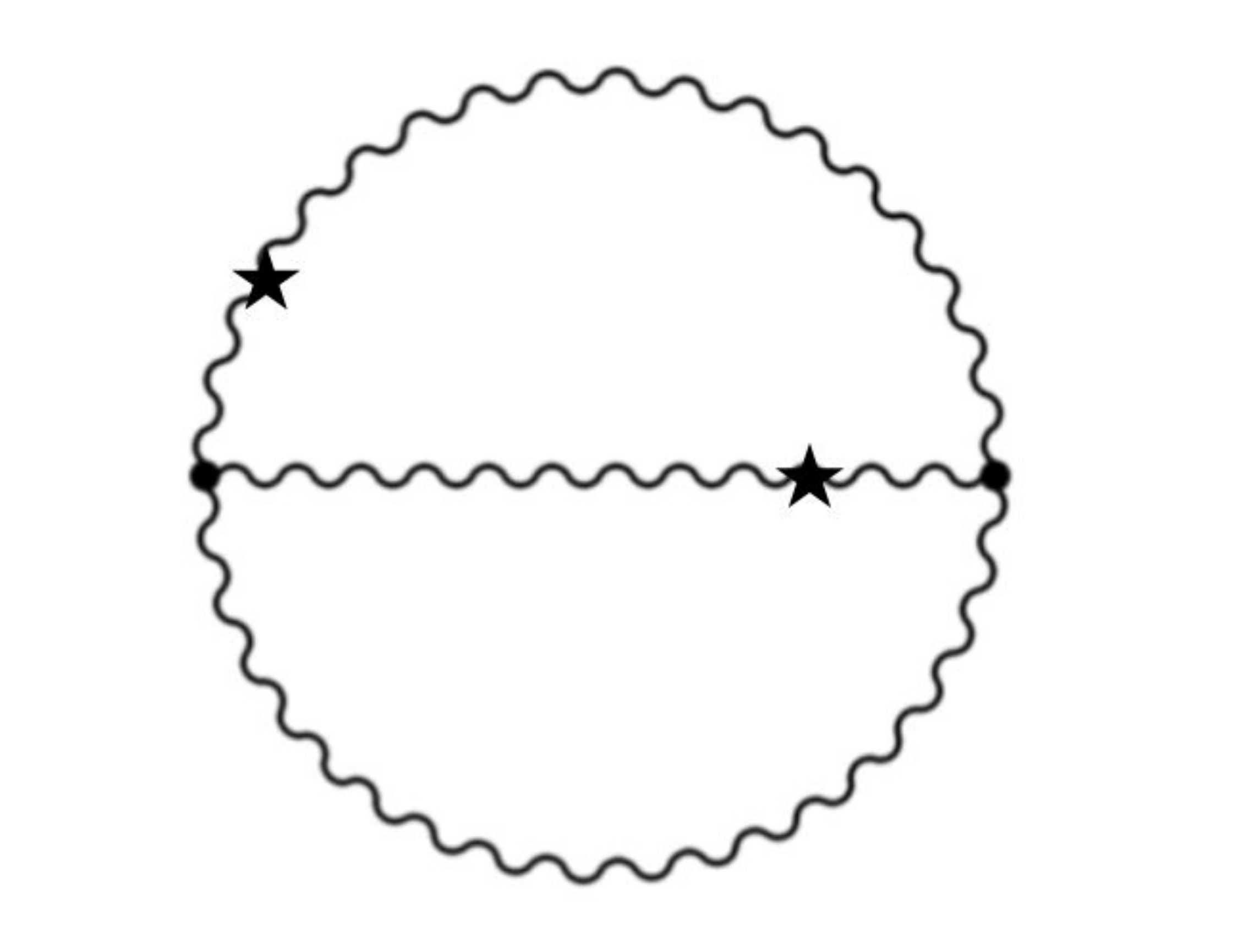}
\\
$A$&$B$&$C$&$D$
\end{tabular}
\end{center}
\caption{Some diagrams with nontrivial symmetry factors (see text).}
\label{fig:sf}
\end{figure}

Let us close this section with a comment on symmetry factors. 
The sources of symmetry factors are the same as in conventional Feynman diagrams:
\begin{enumerate}
\item
Propagators of gauge fields that start and end at a the same vertex produce a factor of $\frac{1}{2}$.
\item
$P$ equivalent propagators give a factor of $\frac{1}{P!}$
\item
$V$ equivalent vertices give a factor of $\frac{1}{V!}$.
\end{enumerate}
The first rule only exists for real fields, i.e.~gauge bosons in the present context. It is the source of the factor of $\frac{1}{2}$ in the last diagram of fig.~\ref{fig:twoloop}.
The other two rules are best illustrated by further examples.
Consider the four diagrams shown in fig.~\ref{fig:sf}.
Diagram $A$ carries a factor of $\frac{1}{2}$ due to rule 2 (the two propagators are interchangeable). Diagram $B$ has two equivalent vertices and two equivalent gauge boson propagators and hence has symmetry factor $\frac{1}{4}$ (the two scalar propagators are non-equivalent since the particle number flows in different directions). 
The presence of the stars  (marking derivatives of fields) typically removes some of the symmetry of the same diagram without stars (as appearing e.g.~in non-covariant perturbation theory). 
Diagram $C$ has two equivalent vertices and two equivalent propagators (the ones without the stars) and hence gets a factor of $\frac{1}{4}$. Finally, the last diagram $D$ has 3 {\em inequivalent} propagators but two equivalent vertices, and hence receives a factor of $\frac{1}{2}$. As always, these 
factors can also be computed by counting Wick contractions, yielding the same results.

\section{Heat kernel coefficients}
\label{sec:b2n}

In this section we briefly review DeWitt's recursive procedure for the calculation of the LHKC's.

Eqns.~(\ref{eq:SE}) and eq.~(\ref{eq:K0def})
 imply the following differential equation for $B(t)$
\be
i\partial_t B(t;x,y)=\left\{[(D^x)^2+X(x)]-it^{-1}(x-y)^\mu D^x_\mu \right\}B(t;x,y)\,.
\ee
Plugging in the expansion eq.~(\ref{eq:hkexp}) yields the recursion 
\be
(n+(x-y)^\mu D^x_\mu)b_{2n}(x,y)=-in[(D^x)^2+X(x)]b_{2(n-1)}(x,y)\,.
\label{eq:cl}
\ee
For $n=0$ this amounts to a differential equation for $b_0$ 
\be
(x-y)^\mu D^x_\mu\, b_0(x,y)=0\,,
\label{eq:b0}
\ee
with the boundary condition $b_0(y,y)=1$ which is a consequence of $K(0;x,y)=K_0(0;x,y)b_0(x,y)=\delta(x-y)$ and $K_0(0;x,y)=\delta(x-y)$. The solution to this equation is the Wilson line
\be
b_0(x,y)=\mathcal P \exp\left(i\int_{y}^x A_\mu ds^\mu\right)\,,
\ee
where $\mathcal P$ denotes path ordering and the line integral is to be taken along the straight line segment connecting $y$ to $x$.

Let us introduce the notation for the coincidence limits, 
\be
\cl{b}(x)\equiv b(x,x)\,.
\label{eq:bracketnotation}
\ee
 Furthermore let us denote covariant differentiation with a semicolon as follows 
\footnote{In the absence of covariant derivatives w.r.t.~the first argument a double semicolon will appears, e.g.~$D^y_\mu b(x,y)=b_{;;\mu}(x,y)$. 
}
\be
b_{;\mu\nu\dots;\rho\sigma\dots}(x,y)\equiv (\cdots D^y_\sigma D^y_\rho)(\cdots D^x_\nu D^x_\mu) b(x,y)\,.
\label{eq:covsemi}
\ee
The coincidence limits for $b_{2n}$ and its covariant derivatives can be computed from repeated covariant differentiation of eq.~(\ref{eq:cl}).
 Let us start with derivatives with respect to the first argument only.
Differentiating $N$ times and using the commutator $[D_\mu,D_\nu]=-iF_{\mu\nu}$, one can derive the generalized recursion relation for the LHKC's
\begin{multline}
(n+N)\cl{b_{2n;\mu_1\cdots \mu_N}}-i \sum_{\substack{L+M=N-2\\i_1<\cdots<i_M\\\ell<j_1<\cdots< j_{L}\\\ell<k}}
F_{\mu_{\ell} \mu_k;\mu_{j_1}\cdots\mu_{j_{L}}}\cl{b_{2n;\mu_{i_1}\cdots \mu_{i_M}}}\\
=-in g^{\nu\rho} \cl{b_{2(n-1);\nu\rho\mu_1\cdots\mu_N }}
-in\sum_{\substack{L+M=N\\i_1<\cdots<i_M\\j_1<\cdots< j_{L}}}
X_{;\mu_{j_1}\cdots\mu_{j_{L}}}\cl{b_{2(n-1);\mu_{i_1}\cdots \mu_{i_M}}}\,.
\end{multline}
 This relation allows one to compute the coincidence limit of the $b_{2n}$ with an arbitrary number $N$ of covariant derivatives in terms of the analogous quantities with smaller $N$ and/or $n$. 
The cases with up to $2n+N=4$, are easily obtained by hand.
\begin{itemize}
\item
Dimension-0 coefficients
\be
\cl{b_0}=1\,.
\ee
\item
Dimension-1 coefficients
\be
\cl{ b_{0;\mu}}=0\,.
\ee
\item
Dimension-2 coefficients
\bea
\cl{ b_{0;\mu\nu}}&=&\frac{i}{2}F_{\mu\nu}\,,\\
\cl{b_2}&=&-iX\,.
\eea
\item
Dimension-3 coefficients
\bea
\cl{ b_{0;\mu\nu\rho}}&=&
	\frac{i}{3}\left(F_{\mu\nu;\rho}	+F_{\mu\rho;\nu} \right)\,,\\
\cl{b_{2;\mu}}&=&\frac{1}{6}F^\nu_{\ \mu;\nu}-\frac{i}{2}X_{;\mu}\,.
\eea
\item
Dimension-4 coefficients
\bea
\cl{b_{0;\mu\nu\rho\sigma}}&=&\frac{i}{4}\left(
	F_{\mu\nu;\rho\sigma}+F_{\mu\rho;\nu\sigma}+F_{\mu\sigma;\nu\rho}  \right)\nn\\
		&&-\frac{1}{8}\left(
			\{F_{\mu\nu},F_{\rho\sigma}\}+\{F_{\mu\rho},F_{\nu\sigma}\}+\{F_{\mu\sigma},F_{\nu\rho}\}
		\right)\,,\\
\cl{b_{2;\mu\nu}}&=& \frac{1}{3}F_{\mu\nu}X+\frac{1}{6}XF_{\mu\nu}-\frac{i}{3}X_{;\mu\nu}\nn\\
				&&-\frac{1}{12}\left(F_{\mu\lambda;\sigma \nu}+F_{\nu\lambda;\sigma\mu}-i\{F_{\mu\lambda},
				F_{\nu\sigma}\}\right)g^{\lambda\sigma}\,,\\
\cl{b_4}&=&-X^2-\frac{1}{3}X_{;\mu}^{\ \ \mu}+\frac{1}{6}F_{\mu\nu}F^{\mu\nu}\,.
\eea
\end{itemize}

The coefficients with derivatives acting on the second argument can be obtained from Synge's rule,
\be
[b_{;\mu}]+[b_{;;\mu}]= [b]_{;\mu}
\ee
 such that we can write yet another recursion
\be
\cl{b_{2n;\mu_1\dots\mu_M;\nu_1\dots\nu_N}}=\cl{b_{2n;\mu_1\dots\mu_M;\nu_1\dots\nu_{N-1}}}_{;\nu_N}
-\cl{b_{2n;\mu_1\dots\mu_M\nu_N;\nu_1\dots\nu_{N-1}}}\,.
\ee
For instance one has
\be
\cl{b_{0;;\mu}}=0\,,
\ee
\be
\cl{b_{0;\mu;\nu}}=-\frac{i}{2}F_{\mu\nu}\qquad \cl{b_{0;;\mu\nu}}=-\frac{i}{2}F_{\mu\nu}\,,
\ee
and
\bea
\cl{b_{0;\mu;\nu\rho}}&=&-\frac{i}{6}\left(F_{\mu\nu;\rho}+F_{\mu\rho;\nu}\right)\,,\\
\cl{b_{0;\mu\nu;\rho}}&=&\frac{i}{6}\left(F_{\rho\mu;\nu}+F_{\rho\nu;\mu}\right)\,,\\
\cl{b_{0;;\mu\nu\rho}}&=&-\frac{i}{3}\left(F_{\mu\nu;\rho}+F_{\mu\rho;\nu}\right)\,,\\
\cl{b_{2;;\mu}}&=&-\frac{1}{6}F^\nu_{\ \mu;\nu}-\frac{i}{2}X_{;\mu}\,.
\eea

The computation of the higher  order terms  becomes somewhat tedious. We have therefore written a small mathematica notebook that computes  coincidence limits of the $b_{2n}$ for arbitary number of covariant derivatives on either argument. The notebook is provided as supplementary material to this article.

In simplifying the results of the LHKC's, a useful relation is the Bianchi identity
\be
F_{\mu\nu;\rho}+F_{\nu\rho;\mu}+F_{\rho\mu;\nu}=0\,.
\label{eq:bianchi}
\ee
Moreover, due to
\be
F_{\rho\sigma;\lambda\mu}-F_{\rho\sigma;\mu\lambda}=-i[F_{\mu\lambda},F_{\rho\sigma}]\,,
\ee
one has
\be
F^{\mu\lambda}_{\ \ ;\mu\lambda}=0\,.
\ee
Finally, elementary Dirac matrix algebra gives some other useful identities,
\be
\gamma^\rho(\gamma^\mu\gamma^\nu F_{\mu\nu})-(\gamma^\mu\gamma^\nu F_{\mu\nu})\gamma^\rho=4\gamma^\mu F^\rho_{\ \mu}\,,
\ee
\be
\gamma^{\mu}\gamma^\nu\gamma^\rho F_{\mu\nu;\rho}=-\gamma^\rho\gamma^\mu\gamma^{\nu} F_{\mu\nu;\rho}=-2\gamma^\mu F^\nu_{\ \mu;\nu}\,,\qquad 
\gamma^\mu\gamma^{\rho}\gamma^\nu F_{\mu\nu;\rho}=0\,,
\label{eq:diracbianchi}
\ee
where in eq.~(\ref{eq:diracbianchi}) we also used the Bianchi identity eq.~(\ref{eq:bianchi}).

\section{Loop integrals}
\label{sec:integrals}

\subsection{Multi-loop momentum integrals}
\label{sec:gaussian}

The most general $L$-loop momentum  integral that we need to evaluate in our formalism is the following simple Gaussian integral:
\be
I(t_i,p_\ell,\xi_i)\equiv\int\left[ \prod_{r=1}^L\frac{d^{d}q_r}{(2\pi)^{d}}\right]
 e^{i\, T_{ij}\, q_i\cdot q_j +2i\, S_{in}\,q_i\cdot p_n+i\, U_{mn}\,p_m\cdot p_n+\xi_i\cdot q_i-im_i^2 t_i}\,.
 \label{eq:I1}
\ee
The matrices $T$, $S$, and $U$ are linear in the Schwinger parameters $t_i$, and in addition $T$ and $U$ are symmetric and   positive definite. 
 We have introduced sources $\xi_i$ in order to account for possible additional terms $q^\mu_i$ in front of the exponential which can be obtained by differentiating with respect to $\xi_i^\mu$.
As mentioned around eq.~(\ref{eq:proptimerot}), in addition to the usual Wick rotation of the momenta we can make the rotation $t_i=-i\tau_i$.
In terms of these Euclidean parameters, we have
\be
I(-i\tau_i,p_n,\xi_i)
 =i^L(4\pi)^{-\frac{dL}{2}}\det(\mathcal T)^{-\frac{d}{2}}
 e^{
   \mathcal U_{mn}p_m\cdot p_n-\mathcal R_{in}\xi_i\cdot p_n -\frac{1}{4} \mathcal T^{-1}_{ij}\xi_i\cdot \xi_j -\tau_im_i^2}\,,
\label{eq:I2}
\ee
where $\mathcal T\equiv iT$, $\mathcal U=i (U-S^TT^{-1}S)$, and $\mathcal R=T^{-1}S$. The matrices $\mathcal T$, $\mathcal U$ and $\mathcal R$ are now purely real in terms of the $\tau_i$, in addition, $\mathcal T$ is positive definite.

After the momentum integration and the expansion in the external momenta as well as the expansion eq.~(\ref{eq:hkexp}),
the remaining $\tau_i$ integrals to be evaluated are of the form 
\be
\int_0^\infty \left(\prod_{i=1}^P \t_i^{a_i-1}d\t_i\right) \frac{\exp(- \t_im_i^2)}{(\det \mathcal T)^{\frac{d}{2}+\gamma}}\,,
\ee
where the $a_i> 0$ and $\gamma\geq 0$ are integers.
Notice that $\det \mathcal T$ is a homogeneous polynomial of degree $L$ in the $\t_i$. 

In the following we will evaluate the general one-loop case and some selected two-loop cases.

\subsection{Schwinger integrals appearing at one loop order}
\label{sec:int1}

At one loop order, there are only bilinear vertices. For $P=P_0+P_m$ propagators (with $P_0$ massless and $P_m$ degenerate massive propagators), there are thus $P$ such vertices.  
The most general one loop graph leads  to the integral
\bea
f(\{a_i\},\{b_j\},c)&\equiv& \int_0^\infty \prod_{i=1}^{P_0} \s_i^{a_i-1}d\s_i\prod_{j=1}^{P_m} \t_j^{b_j-1}d\t_j\frac{e^{-m^2\t }}{(\s+ \t)^c}\\
&=&\frac{\Gamma(a+b-c)}{(m^{2})^{a+b-c}}\
\frac{\Gamma(c-a)}{\Gamma(b)\Gamma(c)}\nn
\prod_{i=1}^{P_0}\Gamma(a_i)\prod_{j=1}^{P_m} \Gamma(b_j)\,,
\eea
where $\s\equiv \sum_{i=1}^{P_0}\s_i$, $\t\equiv \sum_{j=1}^{P_m}\t _j$, $a\equiv \sum_{i=1}^{P_0} a_i$ and $b\equiv \sum_{j=1}^{P_m} b_j$. 

Let us discuss a few special cases.
\begin{itemize}
\item
The zero vertex graph with one heavy field in the loop is always proportional to
\be
f(\{\},n,\tfrac{d}{2})=\frac{\Gamma\left(n-\tfrac{d}{2}\right)}{(m^2)^{n-\frac{d}{2}}}\,,
\ee
where $n$ is the order of the HK expansion that also determines the dimension of the operator, $D=2n$.
\item
A graph with two bilinear vertices and only (degenerate) heavy fields can be written in terms of 
\be
f(\{\},\{b_1,b_2\},c)=\frac{\Gamma(b_1+b_2-c)}{(m^2)^{b_1+b_2-c}}\ \frac{\Gamma(b_1)\Gamma(b_2)}{\Gamma(b_1+b_2)}\,.
\ee
\item
A graph with two bilinear vertices and one heavy and one light field is proportional to
\be
f(a,b,c)=\frac{\Gamma(a+b-c)}{(m^{2})^{a+b-c}}\
\frac{\Gamma(a)\Gamma(c-a)}{\Gamma(c)}\,.
\ee
\end{itemize}

\subsection{Schwinger integrals needed at two loop order}
\label{sec:int2}

Moving on to two loops, let us consider sunset graphs with exactly three propagators (that is, only two trilinear vertices and no bilinear ones).
We will need integrals of the type
\bea
g_{m,m',m''}(a,b,c,e)&=&\int d\t\, d\s\, d\r\  \s^{a-1}\t^{b-1}\r^{c-1} \frac{e^{-\t m^2-\s m'^2-\r m''^2 }}{(\s\t+\s\r+\t\r)^e}\,.
\eea

We distinguish two cases
\be
g_{m,0,0}(a,b,c,d)=\frac{\Gamma(a+b+c-2e)}{
	(m^2)^{a+b+c-2e}}\ 
	\frac{\Gamma(a+c-e)\Gamma(e-a)\Gamma(e-c)}{\Gamma(e)}\,,
\ee
\be
g_{m,m,0}(a,b,c,d)=\frac{\Gamma(a+b+c-2e)}{
	(m^2)^{a+b+c-2e}}\ 
	\frac{\Gamma(c)\Gamma(e-c)}{\Gamma(e)}
 \frac{\Gamma(a+c-e)\Gamma(b+c-e)}{\Gamma(a+b+2c-2e)}\,.
\ee
The integrals for the case of three massive propagators are not expressable in a simple way in terms of $\Gamma$ functions, and we will skip them here.

The other possible two-loop topology is the figure-eight graph. The two loop-momentum integrations are independent and thus result in a product of two  one-loop integrals $f(\{a_i\},\{b_j\},c)$ already evaluated in app.~\ref{sec:int1}.

\acknowledgments
GG wishes to thank the Conselho Nacional de Desenvolvimento Científico e Tecnológico (CNPq) for support under fellowship number 309448/2020-4. KK is supported by Coordenação de Aperfeiçoamento de Pessoal de Nível Superior (CAPES).

\bibliographystyle{JHEP}

\bibliography{literature} 
\end{document}